\documentclass[a4paper,11pt]{article}
\pdfoutput=1 

\usepackage{jheppub} 
\usepackage[T1]{fontenc} 

\usepackage{url}
\usepackage[T1]{fontenc}

\def\ss{\subsection}

\def\ie{\emph{i.e.} }

\def\nt{\notag}

\def\wt{\widetilde}

\def\C{\mathbb{C}}

\def\R{\mathbb{R}}
\def\Z{\mathbb{Z}}

\def\cF{\mathcal{F}}
\def\cG{\mathcal{G}}
\def\cH{\mathcal{H}}
\def\cI{\mathcal{I}}

\def\cM {\mathcal{M}}

\def\cS{\mathcal{S}}
\def\cT{\mathcal{T}}

\def\cV{\mathcal{V}}

\def\bH{\mathbb{H}}

\def\dg {\dagger}
\def\p{\partial}

\def\/{\over}
\def\ov{\over}

\def\rn{\rangle}
\def\ln{\langle}

\def\e{\epsilon}
\def\ve{\varepsilon}
\def\vphi{\varphi}
\def\a{\alpha}

\def\d{\delta}
\def\k{\kappa}
\def\g {\gamma}
\def\la {\lambda}
\def\w {\omega}
\def\z{\zeta}

\def\l{\ell}
\def\mn{{\mu\nu}}

\def\L{\Lambda}
\def\D{\Delta}
\def\G {\Gamma}
\def\Om {\Omega}
\def\S{\Sigma}

\def\ra{\rightarrow}

\def\r{\mathrm}
\def\hc{\text{h.c.}}

\def\_{\hspace{2cm}}
\def\-{\\\notag}
\def\={&=&}

\newcommand\be{\begin{equation}}
\newcommand\ee{\end{equation}}

\newcommand{\bea}{\begin{eqnarray}}
\newcommand{\eea}{\end{eqnarray}}

\newcommand{\bpm}{\begin{pmatrix}}
\newcommand{\epm}{\end{pmatrix}}

\newcommand{\bit}{\begin{itemize}}
\newcommand{\eit}{\end{itemize}}

\newcommand{\ben}{\begin{enumerate}}
\newcommand{\een}{\end{enumerate}}

\newcommand\bsp{\begin{split}}
\newcommand\esp{\end{split}}

\def\le{\left}
\def\ri{\right}

\def\l{\ell}

\def\qq{\qquad}

\def\TT{{T\overline{T}}}

\newcommand{\subf}[2]{
	{\small\begin{tabular}[t]{@{}c@{}}
			#1\\#2
		\end{tabular}}
	}

\title{Quantum cosmology as automorphic dynamics}

\author{Victor Godet}

\affiliation{SISSA and INFN, Via Bonomea 265, 34136 Trieste, Italy}

\emailAdd{vgodet@sissa.it}

\abstract{

We revisit pure quantum cosmology in three dimensions.  The Wheeler-DeWitt equation can be solved perturbatively and the dynamics reduces to a particle on moduli space. Its time evolution is equivalent to the $T\overline{T}$ deformation.
Focusing on spacetimes with torus slices, we show that inflationary cosmologies correspond to particle trajectories in Artin's billiard. The resulting automorphic dynamics is developed both from a first and second quantized perspectives. Our main application is to give an interpretation for the Hartle-Hawking state which is here the analytic continuation of the Maloney-Witten partition function. We obtain its spectral decomposition and an exact representation as an average involving the Möbius function.

}

\begin{document} 
\maketitle
\flushbottom

\bigskip

How quantum mechanics should be applied to the entire universe remains a mystery.  In contrast to their AdS counterparts, cosmological spacetimes have no boundaries and no region where gravity turns off. Thus  it is unclear how to even define gauge-invariant observables  \cite{Witten:2001kn}. This is related to the existence of gravitational constraints \cite{DeWitt:1967yk} which are especially strong for closed universes \cite{Deser:1973zza,Moncrief:1976un,Higuchi:1991tk,Higuchi:1991tm}. Moreover, expanding universes are intrinsically time-dependent so we cannot avoid to study their non-trivial dynamics.

The canonical quantization of gravity \cite{Arnowitt:1962hi,DeWitt:1967yk} does provide a roadmap to define and understand quantum cosmology. There, the Hilbert space is defined as the set of solutions of the Wheeler-DeWitt equation  and the physical operators are those which commute with the constraints. Recent progress has shown that the Wheeler-DeWitt equation can be systematically solved in a large volume limit \cite{Chakraborty:2023yed,Chakraborty:2023los}. In our view, this provides a starting point for the systematic development of quantum cosmology from the canonical perspective. In this work we start this program by going beyond the large volume limit in the simpler context of pure three-dimensional gravity.

Hartle and Hawking defined a wavefunction for the universe using  the gravity path integral \cite{Hartle:1983ai}. Their idea was to sum over all spacetimes with a fixed asymptotic boundary in the future and with a no-boundary condition in the past, resulting in a wavefunctional that automatically solves the Wheeler-DeWitt equation. This proposal has been made more precise in the context of (A)dS/CFT where the Hartle-Hawking state is  realized as an analytically continued CFT partition function \cite{Strominger:2001pn,Maldacena:2002vr,Maldacena:2019cbz}. Although it is unclear whether the Hartle-Hawking state does describe our universe \cite{Maldacena:2024uhs}, it is a very  natural state from the gravitational perspective which deserves to be understood better. This will be one of our goal in this paper.

We will consider pure three-dimensional gravity with a positive cosmological constant and  focus  on spacetimes of the form $\R\times \S$ where $\S$ is a Riemann surface representing space. Riemann surfaces are classified by a moduli space $\cM_g$ for each genus. We will focus on the simplest case of $g=1$ which as we will see is already quite rich.  For $g=1$, the moduli space is two-dimensional and is the modular curve, \ie  the fundamental domain
\be
\cF = \bH/\r{PSL}(2,\Z)
\ee
for the action of $\G=\r{PSL}(2,\Z)$ on the upper half-plane. We will show that the Wheeler-DeWitt equation is equivalent to a Klein-Gordon equation so that quantum cosmology reduces to a driven particle on $\cF$. There is a large literature on quantum cosmology in three dimensions \cite{deserThreedimensionalCosmologicalGravity1984,Witten:1988hc,Witten:1989ip,verlindeConformalFieldTheory1989,Maldacena:1998ih,Witten:2007kt}. The fact that it should reduce to dynamics on moduli space was suggested in many previous works  \cite{martinecSolubleSystemsQuantum1984,hosoyaDimensionalPureGravity1990,carlipNotesDimensionalWheelerDeWitt1994,Moncrief,seriuBackReactionTopological1996,seriuPartitionFunctionDimensional1997} (see \cite{Carlip:2004ba} for a review) but exactly how this equivalence should be realized has remained unclear. See also \cite{Araujo-Regado:2022jpj,Hikida:2022ltr,Chen:2022ozy,Chen:2022xse,Chen:2024vpa} for more recent related works. Our goal in this paper is to explain how this reduction should be understood and  use  it as a starting point to develop quantum cosmology.

The notion of dynamics in a gravitational context has always been puzzling as there is no time in the canonical description. In quantum cosmology, time evolution is ``pure gauge'' but this should not stop us to study the dynamics after we identify an adequate definition of time. York proposed to use one of the canonical variable, the trace of the extrinsic curvature, as a canonical definition of time \cite{York:1972sj}. Here, we will see instead that the total volume of space appears to be the adequate notion of canonical time, as already pointed out in \cite{Chakraborty:2023yed}. This is appropriate for inflationary universes and the resulting dynamics turns out to be quite rich. 

Recently, a particular deformation of two-dimensional CFTs has generated much interest \cite{Zamolodchikov:2004ce,Smirnov:2016lqw}. The so-called $\TT$ deformation gives a solvable irrelevant deformation with many interesting features. In particular, its close relationship with the Wheeler-DeWitt equation has been noticed and discussed in many recent papers \cite{Mazenc:2019cfg,McGough:2016lol,Witten:2022xxp,Araujo-Regado:2022gvw,Araujo-Regado:2022jpj,Chen:2023eic}. We will show that the dynamics governing the particle is equivalent to the  $\TT$ flow. That is, we will see that the $\TT$ deformation, viewed as an evolution equation, does describe inflationary cosmological dynamics.

The classical solutions and the Hartle-Hawking state in the present context were discussed in \cite{Castro:2012gc}. They correspond to inflationary spacetime with torus slices, starting at a singularity and inflating to large volume.  The Hartle-Hawking state is obtained by summing over the solutions which admit a Euclidean continuation with no boundary. These geometries are the set $M_{c,d}$ of $\r{SL}(2,\Z)$ images of the leading geometry $M_{0,1}$, corresponding to all possible ways of filling the torus. Thus the Hartle-Hawking state is the analytic continuation of the Maloney-Witten partition function studied in the AdS$_3$ context \cite{Maloney:2007ud}.

In the particle description, we will see that the Hartle-Hawking state is  the sum over trajectories coming from the infinite cusp. This provides an interpretation of the Hartle-Hawking as a vacuum in the particle theory. The classical solutions $M_{c,d}$ correspond to trajectories in Artin's billiard which leads to a rather intricate dynamics depending on the integers $(c,d)$. We will see that this dynamics can be discretized in terms of the Gauss map.

In some ways, the worldline theory of a particle is similar to quantum cosmology. For the particle, the second-quantized theory gives a more powerful formalism in terms of a quantum field theory in an auxiliary spacetime. As the Wheeler-DeWitt equation does reduce here to the BRST equation of a worldline theory, the second-quantized formulation appears more appropriate. Here it is really a ``third-quantized'' description where we can discuss arbitrary number of universes. The  mode decomposition involves harmonic analysis on $\cF$ and leads to  a natural basis for the Hilbert space in terms of (non-holomorphic) Eisenstein series and Maass cusp forms. Tools from harmonic analysis  will be especially useful. In particular we will obtain the spectral density for the Hartle-Hawking state. A surprising application will be to derive an exact expression for the Maloney-Witten partition function as a $q$-expansion with integer coefficients, with degeneracies given in by the Möbius function. The appearance of such pseudo-random objects from number theory suggests interesting connections between chaotic aspects of automorphic dynamics and gravity.

\section{Cosmology as a particle}

We consider pure Einstein gravity in three dimensions with a positive cosmological constant
\be
S= {1\/16\pi G}\int_M d^3 x\sqrt{-g}\,(R-2\L) - {1\/8\pi G}\int_{\p M} \sqrt{h} K~.
\ee
We focus on spacetimes with topology $\R\times \S$ where $\S$ is a Riemann surface that we take to be a torus. The units are chosen such that $\L=\l_\r{dS}^{-2}=1$. 

In this section, we will argue that this theory is equivalent to a particle on the fundamental domain $\cF= \bH/\r{PSL}(2,\Z)$. The possibility of such a reduction is an old idea \cite{martinecSolubleSystemsQuantum1984,hosoyaDimensionalPureGravity1990,carlipNotesDimensionalWheelerDeWitt1994,Moncrief,seriuBackReactionTopological1996,seriuPartitionFunctionDimensional1997,Carlip:2004ba} which we hope to put on firmer footing. We will explain how this reduction arises in a perturbative expansion and how subleading corrections give  dressing factors associated to Virasoro modes.

\ss{Wheeler-DeWitt reduction}

In canonical gravity, states are functionals $\Psi[g]$ of the spatial metric $g$ that must satisfy the Wheeler-DeWitt equation. For pure three-dimensional gravity, it takes the form
\be
\le[{16\pi G\/\sqrt{g}}\le(  \pi_{ij} \pi^{ij} -\pi^2\ri)  - {1\ov 16\pi G}\sqrt{g}(R-2)\ri]\Psi[g] = 0~.
\ee
Here $\pi^{ij}= -i{\d\/\d g_{ij}}$ is the conjugate momentum of the metric and $R$ is the Ricci curvature on the spatial slice. Writing a solution in the form
\be
\Psi[g] =\r{exp}( i \cF[g])
\ee
we obtain the Hamilton-Jacobi form of the equation
\be\label{HJWDW}
{16\pi G\/\sqrt{g}}\le(  (\pi_{ij}\cF)(\pi^{ij}\cF)  -(\pi\cF)^2\ri)  + {1\ov 16\pi G}\sqrt{g}(R-2) =0~.
\ee
The Hamilton-Jacobi form is convenient because it contains only the principal part of the  equation and ignores contact terms. Such terms are ambiguous as they depend on the normal ordering prescription. They can be modified by multiplying the wavefunction by counterterms so we will ignore them for now.

The spatial metric is two-dimensional and hence can be put in the conformally flat form
\be\label{metricWDW}
ds^2= \Om(u,v){|du+\tau dv|^2\/y},\qq u\sim u+2\pi,\,\,\, v\sim v+2\pi~.
\ee
Hence the degrees of freedom are the moduli $\tau=x+iy$ and the Weyl factor $\Om(u,v)$. 

Of particular importance will be the total volume
\be
T={1\/4\pi^2}\int d^2u \,\sqrt{g}  = {1\/4\pi^2} \int dudv\,\Om(u,v)
\ee
as this will be our definition of canonical time. It will be particularly useful to take the limit $T\to+\infty$, which corresponds to a large volume regime. Taking $T$ to be the canonical definition of time is appropriate for inflationary universes.

We can impose the  momentum constraints by restricting to wavefunctionals that are invariant under spatial diffeomorphisms
\be
\Psi[g^{(\xi)}]=\Psi[g]~.
\ee
The Hamilton-Jacobi equation then takes the form
\be
(4\pi^2\d_\Om \cF)^2 -{y^2\/\Om^2}\le( (\p_x \cF)^2 +(\p_y \cF)^2\ri) ={\pi^2\/4G^2}(1-\tfrac12 R)~.
\ee
To obtain this we have  performed separation of variables. The variation of the metric \eqref{metricWDW} is given as
\be
\d g_{ij} = \Om^{-1} g_{ij}\d\Om +  \p_x g_{ij} dx+\p_y g_{ij}dy
\ee
and leads to
\be
{\d\Psi\/\d\Om} = \Om^{-1}g_{ij}{\d\Psi\/\d g_{ij}},\qq {\p \Psi\/\p x} = \int du dv\,\p_x g_{ij}{\d\Psi\/\d g_{ij}},\qq {\p \Psi\/\p y} = \int du dv\,\p_y g_{ij}{\d\Psi\/\d g_{ij}}~,
\ee
which can be inverted as
\be
{\d\Psi\/\d g_{ij}} = {1\/2}g^{ij}\Om{\d\Psi\/\d\Om} - {1\/4\pi^2}{y^2\/2}\Big( \p_x g^{ij} {\p\Psi\/\p x} + \p_y g^{ij} {\p\Psi\/\p y}\Big)~.
\ee
This follows from the orthogonality relations
\be
g^{ij}\p_x g_{ij} = g^{ij}\p_y g_{ij}= \p_x g^{ij}\p_y g_{ij}=0,\qq \p_x g^{ij}\p_x g_{ij} = \p_y g^{ij} \p_y g_{ij} = -{2\/y^2}~.
\ee
Thus we obtain ordinary derivatives in the moduli $x$ and $y$. However we still have a functional derivative in $\Om(u,v)$.

To make further progress, we consider a perturbative expansion around a flat metric where we set $\Om(u,v)=T$ so we take the spatial metric to be
\be\label{perturbedMetric}
ds^2 = T {|du+\tau dv|^2\/y} (1+\ve\, \vphi(u,v)+O(\ve^2))
\ee
where $\ve$ is a perturbative parameter. We will also take a perturbative ansatz for the wavefunction of the form
\be
\Psi[g]=e^{i\cF[g]},\qq \cF[g]= S(T,x,y) + \ve \,\cV[g]+O(\ve^2)~.
\ee
That is, we demand that at leading order in $\ve$, $\cF[g]$ depends  only on the total volume $T={1\/4\pi^2}\int d^2 u \sqrt{g}$ and the moduli $x,y$. This is similar to a WKB approximation. Here, the perturbative expansion could be the large volume expansion or the expansion in Newton's constant.

In this approximation, the action of $\d_\Om$ is simply
\be
{\d\/\d\Om(u,v)}S(T,x,y) = {1\/4\pi^2} \p_T S 
\ee
and reduces to a constant. The Ricci scalar is $R=O(\ve)$ so the  Wheeler-DeWitt equation reduces to an ordinary differential equation
\be
(\p_T S)^2 -{y^2\/T^2}\le( (\p_x S)^2 +(\p_y S)^2\ri) -{\pi^2\/4G^2} =O(\ve)~.
\ee
This is the Hamilton-Jacobi form of the Klein-Gordon equation for a particle
\be
(\Box+M^2)\Psi=0,\qq M ={\pi\/2G}~.
\ee
living on the auxiliary spacetime 
\be\label{ds2Auxsec1}
ds^2_\text{Aux} = dT^2 - T^2{dx^2+dy^2\/y^2}~.
\ee
The mass $M$ of the particle is the cosmological scale $\l_\r{dS}$ in Planck units. Note that it is actually equal to the dS$_3$ entropy.

It is instructive to consider the first subleading correction. At $O(\ve)$ we obtain
\be\label{oneloopWDW}
{\d\/\d\Om} \,\cV = {1\/32\pi G}R + {G\/2\pi^3}{y^2\/T^2} ((\p_x S)^2+(\p_y S)^2) \vphi
\ee
The first term shows that  $e^{i\cV[g]}$ transforms as a CFT partition function with central charge $c={3i\/2G}$. The second term is subleading in a small $G$ or large $T$ expansion. Thus we see that the one-loop correction provides a dressing factor for the particle. As discussed below, we will see that it corresponds to the contribution of the Virasoro modes.

\ss{Time evolution from $\TT$}

The solutions of the Wheeler-DeWitt equation at late time have the same symmetries as CFT partition functions  \cite{Chakraborty:2023yed}. In three dimensions, we have seen that the leading order dynamics reduces to that of a Klein-Gordon particle. Here we will see that this dynamics is equivalent to applying the $\TT$ deformation to the wavefunctional, viewed as a CFT partition function. This could have been anticipated from the known relationship between the $\TT$ deformation and the Wheeler-DeWitt equation \cite{McGough:2016lol,Mazenc:2019cfg,Witten:2022xxp,Araujo-Regado:2022gvw,Araujo-Regado:2022jpj,Chen:2023eic}. 

Let us consider again the Hamilton-Jacobi form \eqref{HJWDW} of the Wheeler-DeWitt equation. After redefining $\cF$ with a counterterm
\be
\cF[g] = S_\r{ct}[g]+ \cF_0[g],\qq S_\r{ct}[g]= {1\/8\pi G} \int d^2 x\sqrt{g}~.
\ee
the finite piece $\cF_0$ satisfies the equation
\be
2 i (\pi \cF_0) - {1\/2\k^2}\sqrt{g} R = {16\pi G\/\sqrt{g}} \le( (\pi_{ij}\cF_0)(\pi^{ij}\cF_0) - (\pi\cF_0)^2 \ri)~.
\ee
At leading order  we can ignore the RHS and this is the Weyl anomaly equation for a CFT of central charge $c={3i\/2G}$.

The other terms can be generated if we deformed the CFT by the $\TT$ operator
\be
S_\r{QFT} = S_\r{CFT}+ \int d^2 u\,\la\, T\overline{T}~,
\ee
with deformation parameter $\la$. Here the definition of the $\TT$ operator is
\be
\TT  = {1\/8}\sqrt{g} (T_{ij}T^{ij}-(g_{ij}T^{ij})^2)~.
\ee
 The Weyl anomaly equation is then modified to
\bea
\ln g^{ij}T_{ij}\rn \= {c\/24\pi} R+ 2\la \ln T\overline{T}\rn~.
\eea
Zamolodchikov showed that the expectation value of $\TT$ in a flat metric takes the form \cite{Zamolodchikov:2004ce}
\be
\ln T\overline{T}\rn = \ln T\rn \ln \overline{T} \rn -{1\/16}\ln g^{ij}T_{ij}\rn^2~.
\ee
The stress tensor acts on CFT partition functions  as $T^{ij}=  -{2\/\sqrt{g}}{\d\/\d g_{ij}}$ which is essentially the conjugate momentum $
\pi^{ij}$. As a result, we see that if we consider the functional 
\be
\Psi[g] = e^{i S_\r{ct}[g]}Z_\r{\TT}[g]~,
\ee
where $Z_\r{\TT}[g]$ is the $\TT$-deformed CFT partition function with central charge $c$ and deformation parameter $\la$, we have that $\Psi$ solves the Wheeler-DeWitt equation with
\be\label{clambda}
c={3i\/2G},\qq \la = {16i\pi G\/\sqrt{g}}={4i G y\/\pi T}~.
\ee
Thus we see that the deformation parameter $\la$ is essentially the inverse volume of the metric. Here the matching involves using  holomorphic coordinates $w = {1\/2\pi}(u+\tau v)$ in the metric so that its determinant is $\sqrt{g}= {T\/4\pi^2 y}$ as we have
\be\label{torusMetric}
ds^2 = T {|du+\tau dv|^2\/y} = {T\/4\pi^2 y} |dw|^2
\ee
which is responsible for the additional factor of $y$. Here we have restricted to flat metrics so this only captures the leading order solution as described in the previous section.

On the torus, the flow of $\TT$-deformed partition functions have been extensively studied \cite{Aharony:2018bad,Cardy:2018sdv,Datta:2018thy}. CFT partition functions on the torus are modular functions and this leads to a notion of $\TT$-deformed modular forms \cite{Cardy:2022mhn}. It was shown that $\la$ transforms under $\r{SL}(2,\Z)$ in the same way as  $y=\r{Im}\,\tau$. As a result the combination 
\be
T = {4i G\/\pi} {y\/\la}
\ee
is modular invariant. This is expected because $\r{SL}(2,\Z)$ is a gauge symmetry here, as the diffeomorphism $\tau \ra \g\tau$ can be compensated by a diffeomorphism $(u,v)=(u,v).\g$ without affecting $T$.

It was also shown that  $\TT$ flow can be solved exactly and written as a kernel integral \cite{Freidel:2008sh,Datta:2021kha,He:2024pbp,Iliesiu:2020zld}
\be\label{TTkernel}
Z_\la(x,y) = - {y \/\pi\la}\int {dx'dy'\/y'^2} \r{exp}\le( {|\tau-\tau'|^2 \/ \la y'} \ri)Z_0(x',y')~,
\ee
where we write $\tau=x+iy,\tau'=x'+iy'$. The wavefunctional then takes the form
\be\label{psieff}
\Psi(T,x,y)= {1\/T}\,e^{i S_\r{ct}[g]} Z_\la(x,y)~
\ee
in terms of the $\TT$-deformed partition function $Z_\la(x,y)$. Note that here we are considering here the reduced wavefunctional, evaluated on the flat metric \eqref{torusMetric}. In this expression the value of $\la$ should be set to \eqref{clambda} and the extra factor of $T^{-1}$ can be viewed as a normal ordering prescription.

Then we can verify that the wavefunction \eqref{psieff} defined from the kernel \eqref{TTkernel} does satisfies the Klein-Gordon equation
\be
(\Box+M^2)\Psi=0
\ee
on the auxiliary space \eqref{ds2Auxsec1}. This shows that the dynamics of the particle is equivalent to the $\TT$ deformation. This gives an explicit realization of the idea that the emergence of time in cosmology corresponds to a renormalization group flow.

Note that the analysis here is only valid for flat spatial metrics so it only captures the leading order approximation, as described in the previous section. One could hope that  $\TT$ also gives the general solution. This was suggested in \cite{McGough:2016lol} but it would require understanding the $\TT$ deformation on curved spaces as the deformation parameter $\la$ is essentially the Weyl factor of the metric, see \cite{Mazenc:2019cfg} for a discussion about this.

\ss{Path integral reduction}

It is also illuminating to consider the reduction from a path integral perspective where it is simply Kaluza-Klein reduction on the torus. We can restrict to metrics of the form
\be\label{canMetric}
ds^2 = -N(t,u,v)^2 dt^2+ \Om(t,u,v){|du+\tau(t) d v|^2\/y(t)},\qq u\sim u+2\pi,\,\,\,v\sim v+2\pi~.
\ee
Here we have used a redefinition of $t$ to cancel the components $g_{tu}$ and $g_{tv}$ of the metric. Then with a spatial diffeomorphism we can always bring the spatial metric  to be a Weyl factor $\Om(t,u,v)$ times a flat metric. Thus the moduli $\tau(t)=x(t)+i y(t)$ can be taken to be a function of $t$ only. 

Evaluating the on-shell action gives
\be
S= -{1\/16\pi G} \int dt du dv \le[ {1\/2 N \Om} \le(\dot{\Om}^2-\Om^2 {\dot{x}^2+\dot{y}^2\/y^2}\ri)+(2-R) N T\ri]~,
\ee
where the  Gibbons-Hawking-York boundary term implements an integration by part in time. We can then integrate over $N$ which give
\be
S= -{1\/8\pi G}\int dt du  dv \sqrt{1-\tfrac12 R}\,\sqrt{\dot{\Om}^2- \Om^2 \tfrac{\dot{x}^2+\dot{y}^2}{y^2}}~.
\ee
This is almost the action of a particle but we still have spatial dependence $\Om=\Om(t,u,v)$ and a term involving the Ricci curvature of the slice $R=-\Box_{T^2}\log\Om$. As a result this is a complicated theory of the Weyl factor.

The particle is obtained when $\Om$ is taken to be approximately spatially independent. To investigate this we can use a  decomposition
\be
\Om(t,u,v) = T(t) \,e^{\vphi(t,u,v)},\qq \int du dv\, e^{\vphi}=1~,
\ee
where to avoid redundancies we have imposed that $\vphi$ does not change the total volume.
We then obtain
\be
S = -{1\/8\pi G} \int dt du dv\,\sqrt{1+\tfrac12\Box_{T^2}\vphi} \,\sqrt{ (\dot{T}+T \dot\vphi(t,u,v))^2 - {T^2}\tfrac{\dot{x}^2+\dot{y}^2}{y^2}}~.
\ee
At leading order we set $\vphi=0$ and  we recover the particle 
\be
S=  -M\int dt \sqrt{\dot{T}^2- T^2 \tfrac{\dot{x}^2+\dot{y}^2}{y^2}}~
\ee
after integrating over space. At higher orders, there is a non-trivial coupling between the particle and the Weyl factor of the metric. This provides a dressing factor for the particle. At late time, we can take the metric to be flat and the only contribution will come from Virasoro modes.

Note that  by canonically quantizing this theory we obtain the Hamiltonian
\be
H = \int du dv\,N T\le(  8\pi G\Big({-}\pi_\Om^2 + {y^2\/T^2}(\pi_x^2+\pi_y^2)\Big) + {1\/16\pi G} (2-R)\ri)
\ee
which reproduces   the Wheeler-DeWitt equation. This makes it clear that at leading order, the Wheeler-DeWitt equation  is the same as the BRST constraint of a worldline theory
\be
Q_\r{BRST} = 0~,
\ee
which is equivalent to the Klein-Gordon equation in the auxiliary spacetime. From this perspective it is natural to consider the second-quantized formulation.

\ss{One-loop corrections}

At subleading order, the correction to the particle is given by \eqref{oneloopWDW}. The second term in the RHS of that equation is subleading so this gives the Weyl anomaly equation for a CFT partition function of central charge $c={3i\/2G}$. As a result we can write a solution in terms of the Polyakov action \cite{Polyakov:1987zb}
\be
\cV[g] = {1\/64\pi G} \int d^2x \sqrt{g} \,R\,{1\/\Box }R~.
\ee
If we evaluate this on a flat metric, this will give rise to the geometric action for Virasoro modes  which computes the Virasoro character on the torus. This can be obtained by decomposing the Polyakov action into different contributions \cite{verlindeConformalFieldTheory1989,Nguyen:2021pdz,deBoer:2023lrd}. A simple way to see this here is to write the Polyakov action as
\be
e^{i\cV[g]} = \int D\vphi \,e^{iS_L[\vphi]},\qq S_L[\vphi] = {1\/32\pi G}\int d^2 x\sqrt{g}\,(\tfrac12(\p\vphi)^2+\vphi R)~,
\ee
and setting the metric to be flat simply gives a contribution of the determinant of the Laplacian on the torus
\be
{1\/\sqrt{\r{det}(\Box_{T^2})} }=  {1\/\sqrt{y}|\eta(\tau)|^2}~,
\ee 
which is equal to the Virasoro character, see for example \cite{quineZetaRegularizedProducts1993,Giombi:2008vd}. Thus we see that the late time wavefunctional is
\be\label{Psiphi}
\Psi(T,x,y) = {1\/\sqrt{y}|\eta(\tau)|^2} \,\phi(T,x,y)
\ee
where $\phi$ is the Klein-Gordon field discussed above. This takes the form of a scalar field dressed by a Virasoro factor. This is the exact late time expression since the path integral is one-loop exact in this limit. If we view $\Psi$ as a CFT partition function, we see that the Klein-Gordon field $\phi$ captures the primary partition function obtained by removing the Virasoro descendants.

Note that the Polyakov action also arises in a path integral representation of the late-time wavefunction. There it appears explicitly as an integral over residual gauge modes as explained in  \cite{Carlip:2005tz,Nguyen:2021pdz}.

It's interesting to interpret this factor as coming from a Virasoro average. The point is that, at late time, we can fix the diffeomorphism and Weyl gauge symmetries by setting the metric to be flat. But this leaves a residual gauge group consisting in the Virasoro modes. As the wavefunction transforms as a CFT partition function, upon changing the Weyl factor we get
\be
\Psi[e^{\vphi}\g(T,x,y)] = e^{iS_L[\vphi]} \Psi[\g(T,x,y)]~.
\ee
 The Virasoro average corresponds to integrating over $\vphi$ which does reproduce the Polyakov action. 

 More generally we expect that what appears here is the geometric action for the Virasoro group \cite{Witten:1987ty,Alekseev:1988ce,Alekseev:1990mp,Coussaert:1995zp}. It was emphasized recently \cite{Cotler:2018zff,Nguyen:2022xsw} this is precisely the theory which computes Virasoro blocks. This must be viewed as a proper way to integrate over the Virasoro group in the same vein as the treatment of the de Sitter isometry group in perturbative de Sitter quantum gravity where the perturbative solutions to the constraints can be realized as group-averaged wavefunctionals \cite{Higuchi:1991tm,Marolf:2008it,Chandrasekaran:2022cip,Chakraborty:2023yed,Chakraborty:2023los}. 

\ss{Inner product}

The natural inner product for the particle is the Klein-Gordon inner product
\be
(\phi,\phi) =i \int {dx dy\/y^2}(\phi T^2\p_T\phi^\ast- \phi T^2\p_T\phi^\ast)~.
\ee
Note that this inner product is essentially fixed by consistency with the Hamiltonian constraint. It is the inner product that makes the time evolution  unitary.

At late time, the finite piece of the functional can be defined as the limit
\be
Z(x,y) = \lim_{T\to+\infty}  T e^{iMT} \Psi(T,x,y) 
\ee
where the $T e^{i MT}$ are the necessary counterterms to obtain a finite limit. Then $Z$  can be interpreted as a CFT partition function.

From the relationship  \eqref{Psiphi} we obtain the  late time behavior
\be
\phi(T,x,y) \sim {1\/ T}\,e^{-iMT} \sqrt{y}|\eta|^2 Z(x,y) ~.
\ee
The Klein-Gordon inner product is time-invariant so can be computed at any time. In particular at late time we get
\be\label{lateTimeKG}
(\phi,\phi) = \lim_{T\to+\infty}(\phi,\phi) = \int{dx dy\/y^2} |Z(x,y)|^2  y|\eta(\tau)|^4~.
\ee
This is precisely the inner product that we expect on $\cI^+$. It takes the form \cite{Chakraborty:2023los} 
\bea\label{finalInner}
(\Psi,\Psi) \=\int {Dg\/\r{vol}(\text{diff}\times \text{Weyl})} |Z[g]|^2 
\eea
and the gauge can be fixed in a way familiar from string theory \cite{polchinskiEvaluationOneLoop1986}. The integral over metrics splits into an integral over the diffeomorphism and Weyl groups and an integral over moduli space
\be
Dg = D\xi D\vphi \, {dxdy\/y^2} Z_{bc}(x,y),\qq Z_{bc}(x,y)=y |\eta(\tau)|^4
\ee
with a Jacobian factor equal to the partition function of the $bc$ ghost CFT with $c=-26$. This reproduces the formula \eqref{lateTimeKG} which shows that the two prescriptions for the inner product agree. The upshot is that the one-loop Virasoro prefactor in the wavefunction precisely cancels the factor of $Z_{bc}$ when computing an inner product on $\cI^+$, resulting in the Klein-Gordon inner product for the particle.

For the definition of the inner product \eqref{finalInner} to be consistent we need that the central charge of the CFT partition function $Z$ to be
\be\label{cLate}
c= 13+ {3i\/2G} + iO(1),\qq \la\in \R
\ee
so that the central charge of $|Z|^2$ cancels the  contribution $c=-26$ from the ghosts. Additional corrections to $c$ are allowed as long as they  are pure imaginary. Note that the correction by $13$ can be derived from a one-loop computation in the path integral \cite{Giombi:2008vd,Cotler:2018zff}.

\section{First quantization}\label{sec:First}

We have seen that pure quantum cosmology in three dimensions reduces at leading order to a particle on moduli space described by the action 
\be
S= -M \int dt\,\sqrt{\dot{T}^2- T^2\tfrac{\dot{x}^2+\dot{y}^2}{y^2}}~.
\ee
The equations of motion are simply the geodesic equations on the auxiliary spacetime
\be\label{Auxmetric}
ds^2_\text{Aux} = dT^2-T^2{dx^2+dy^2\/y^2}~.
\ee
Here $\tau=x+iy$ covers the fundamental domain $\cF=\bH/\r{PSL}(2,\Z)$. It will be simpler to consider the theory as a particle in the Teichmüller space $\bH$ and view $\r{PSL}(2,\Z)$ as a gauge symmetry. Recall that this gauge symmetry corresponds to the fact that any transformation of $\tau\ra \g\tau$ can be compensated by a spatial diffeomorphism $(v,u)\ra (v,u).\g$ which must be in $\r{SL}(2,\Z)$ to preserve the periodicity conditions $u\sim u+2\pi,v\sim v+2\pi$.

\ss{$\r{SL}(2,\R)$ symmetry}

The theory can be written in Polyakov form
\be
S =- {M\/2}\int dt \le(e^{-1}\Big(\dot{T}^2-T^2 {\dot{x}^2+\dot{y}^2\/y^2}\Big) +e\ri)
\ee
where $e(t)$ is a Lagrange multiplier which is set  to
\be
e(t)=\sqrt{\dot{T}^2-T^2\tfrac{{\dot{x}^2+\dot{y}^2}}{{y^2}}}~,
\ee
and we could choose for example $e=1$ in the affine parametrization of the trajectory. The relation with the lapse $N$ is $e = 2NT$.

The auxiliary metric \eqref{auxMetricSecond} is flat so it corresponds to a patch of three-dimensional Minkowski space
\be
ds^2_\text{Aux} = dZ^2 - dX^2 - dY^2
\ee
where the mapping between the coordinates is
\be\label{TxyToXYZ}
T^2= Z^2- X^2-Y^2,\qq x = {X\/Z-Y} ,\qq y^2 = {Z^2-X^2-Y^2\/(Z-Y)^2}~,
\ee
and whose inverse is
\be\label{XYZtoTxy}
X= {Tx\/y},\qq Y= {T(x^2+y^2-1)\/2y},\qq Z ={T(x^2+y^2+1)\/2y}~.
\ee
The theory possesses a (local) $\r{SL}(2,\R)$ symmetry coming from the corresponding symmetry on $\bH$. In the embedding space the $\r{SL}(2,\R)$ symmetry is realized as the $\r{SO}(1,2)$ symmetry of $\R^{1,2}$. In particular  the variable $T$ is invariant under $\r{SL}(2,\R)$.

The coordinates $(T,x,y)$ provide  a slicing of the future wedge of $\R^{1,2}$ into hyperboloids with $T=\r{const}$. In particular we see that the $T=0$ surface corresponding to universes of zero size is the light-cone in $\R^{1,2}$.

The $\r{SL}(2,\R)$ symmetry is generated by the Killing vectors
\be
\xi_{-1} = \p_x,\qq \xi_0 = x\p_x+y\p_y,\qq \xi_{+1}= (x^2-y^2)\p_x+2xy\p_y
\ee
and the corresponding Noether charges associated with a trajectory $t\mapsto(T(t),x(t),y(t))$ 
take the form
\be
Q_{-1} = M{T^2\/ e y^2}\dot{x},\qq Q_0 = M{T^2\/e y^2}(x\dot{x}+y\dot{y}),\qq Q_1 =M {T^2\/e y^2}((x^2-y^2)\dot{x}+2xy\dot{y})~.
\ee
Of particular importance is the  Casimir 
\be
C_2 = {1\/2}(Q_{-1}Q_1 +Q_{1}Q_{-1})-Q_0^2= -M^2{T^4\/e^2 y^2}(\dot{x}^2+\dot{y}^2)~.
\ee
These charges are conserved for any solution of the equations of motion. In the Hamiltonian formalism, they become operators
\be
Q_{-1} = \pi_x,\qq Q_{0}=  (x\pi_x+y\pi_y),\qq  Q_1 = (x^2-y^2)\pi_x + 2 xy \pi_y
\ee
where $\pi_x=-i\p_x,\pi_y=-i\p_y$ are the conjugate momenta. We see that they implement the $\r{SL}(2,\R)$ symmetry on the wavefunctions. The Casimir is the Laplacian
\be
C_2 = -y^2 (\pi_x^2+\pi_y^2) = -\D~,
\ee
where in our conventions $\D= -y^2(\p_x^2+\p_y^2) \geq 0$. 

In particular we see that the Casimir is negative
\be
C_2<0
\ee
so the relevant $\r{SL}(2,\R)$ representations correspond to the continuous series.
In Minkowski coordinates, the Casimir depends on the initial point and direction of the trajectory
\be
C_2 = \vec{X}_0^2 - {(\vec{P}\cdot\vec{X}_0)^2\/ \vec{P}^2}~.
\ee
We have $\vec{X}_0^2=0$ as we consider trajectories starting on the light-cone $T=0$. As we need $\vec{P}^2>0$ to be in the future light-cone, the Casimir is indeed negative.

Given a trajectory solution, the projected trajectory $t\mapsto (x(t),y(t))$ is a geodesic of $\bH$ which satisfies the algebraic  equation
\be
(Q_{-1}x(t)-Q_0)^2 +Q_{-1}^2 y(t)^2 + C_2=0~.
\ee
For $Q_{-1}\neq0$, this corresponds to circles centered on the real line while for $Q_{-1}=0$ this gives vertical lines.

\ss{Trajectories as spacetimes}

A particle trajectory 
\be
t\mapsto (T(t),x(t),y(t))
\ee
is associated with  a cosmological spacetime
\be\label{FGmetric}
ds^2=  -{e(t)^2\/4T(t)^2} dt^2 + {T(t)\/y(t)} \,(du+x(t) dv)^2 + T(t)\,y(t) \,dv^2~.
\ee
In particular the equations of motion for the particle are equivalent to Einstein's equation for this spacetime 
\be\label{Einsteineq}
R_\mn -{1\/2} Rg_\mn + g_\mn=0~.
\ee
This follows from the reduction explained in the previous section and can be checked explicitly here.

Although the flat space realization makes affine trajectories particularly simple to study, we will write the trajectories in Fefferman-Graham gauge corresponding to the choice of parametrization
\be
{e(t)\/ T(t)} = {1\/t}~.
\ee
We will fix a torus with moduli $\tau=x+iy$ and consider trajectories which asymptote to it
\be
\lim_{t\to+\infty}(T(t),x(t),y(t)) = (+\infty,x,y)~.
\ee
The leading trajectory is a vertical trajectory and we can solve the equations of motion to obtain
\be
T(t) = t-{t_0^2\/t},\qq x(t)=x,\qq y(t) = {t+t_0\/t-t_0}y~.
\ee
This trajectory is defined for $t>t_0$ where we take $t_0>0$. We see that 
\be
\lim_{t\to+\infty}(T(t),x(t),y(t)) = (+\infty,x,y),\qq \lim_{t\to t_0}\,(T(t),x(t),y(t)) = (0,x,+\infty)~,
\ee
so the trajectory starts at the cusp $\tau=i\infty$ and goes to the final point $(x,y)$. This will be the trajectory corresponding to $M_{0,1}$.

The other trajectories are represented by portions of circles centered on the real line.  The solutions can be parametrized by the final point $(x,y)$ and the initial point $(x_0,0)$ in addition to the parameter $t_0$. The solution takes the form
\bea\nt
T(t)\= t-{t_0^2\/t}~,\\\label{trajGEN}
x(t)\=x + {4t_0 t(x_0-x)y^2\/(t-t_0)^2 (x-x_0)^2 + (t+t_0)^2 y^2}~,\-
y(t)\= y {(t^2-t_0^2) (y^2+(x-x_0)^2)\/(t-t_0)^2(x-x_0)^2+(t+t_0)^2 y^2}~.
\eea
This corresponds to the general trajectory such that
\be
\lim_{t\to+\infty} (T(t),x(t),y(t)) = (+\infty, x,y),\qq \lim_{t\to t_0} \,(T(t),x(t),y(t)) = (0,x_0,0)~.
\ee
In both cases the Casimir is given by
\be\label{C2tot0}
C_2 = -4 M^2 t_0^2 ~,
\ee
and the on-shell action can be computed as
\be
S(T) =-  M \int_{t_0}^{t} dt \Big(1-{t_0^2\/t^2}\Big) = -M T + 2 M t_0 + O(T^{-1})~
\ee
in the limit $T\to+\infty$. The first term is the familiar counterterm and the finite part of the action is given by the Casimir
\be
S_\text{finite}=  2M t_0 = \sqrt{|C_2|}~.
\ee
These trajectories correspond to inflationary spacetimes with torus slices. They were discussed in the present context in \cite{Castro:2012gc} and can be viewed as quotient spacetimes dS$_3/\Z$.

\ss{The no-boundary condition}\label{sec:noboundary}

For now we have discussed Lorentzian spacetimes corresponding to inflationary cosmologies. The limit $T\to0$ is singular and one expects that the semi-classical approximation of gravity breaks down here.

The Hartle-Hawking proposal suggests that one should consider complexified geometries and impose a no-boundary condition. This leads to a well-defined semi-classical gravity path integral, at least at the level of a sum over saddle-points. To obtain this we require that as $T\to0$, the geometry caps off smoothly after an appropriate analytic continuation.

For the trajectory $M_{0,1}$ the metric takes the explicit form
\be
ds^2=  -{dt^2\/4t^2} + {1\/t}\le( {(t-t_0)^2\/y} (du+x dv)^2 + (t+t_0)^2 y\, dv^2 \ri)
\ee
The value $T=0$ corresponds to $t=t_0$ and we see that the geometry cannot be made smooth because the signs of the first and second terms are opposite. 

The proper way to analytically continue such geometries was suggested by Maldacena  \cite{Maldacena:2002vr,Maldacena:2019cbz} and discussed by Castro-Maloney in the present context \cite{Castro:2012gc}. The usual Poincaré coordinate is  $t=z^2$  for a metric normalized as $ds^2 = -{dz^2\/z^2}+\dots$. The prescription for analytic continuation is  to take $z\ra i z$ to convert the asymptotically de Sitter metric into a minus Euclidean AdS metric. Here we see that this  corresponds to simply changing the sign of $t$. As a result we will consider the Euclidean metric
\be
ds^2_\r{Euc}=  -{dt^2\/4t^2} - {1\/t}\le( {(t-t_0)^2\/y} (du+x dv)^2 + (t+t_0)^2 y\, dv^2 \ri)
\ee
where we have also redefined the sign of $t_0$ and we take $t_0>0$ for definiteness. This metric is defined for $t>t_0$ and the first circle will shrink smoothly at $t=t_0$. It still satisfies the Einstein equation \eqref{Einsteineq} and can be interpreted as minus the metric of the Euclidean BTZ black hole which is just  a filled torus $S^1\times D_2$. 

Expanding this metric around $t=t_0$ using $t=t_0+\la r$, we focus on the $du$ circle by taking a slice $v=\r{const}$, and the metric  becomes
\be
ds^2_\r{Euc}=-{1\/4t_0^2\la^2} \Big(dr^2 + {4r^2 t_0\/y} du^2\Big)~.
\ee
From this we see that the circle shrinks smoothly at $r=0$ if we impose the relation
\be\label{t0M01}
t_0 = {y\/4}~.
\ee
In particular the on-shell action then becomes
\be
S_\text{finite}= {M\/2}y~,
\ee
which is the expected on-shell action of the $M_{0,1}$ spacetime. Thus the no-boundary condition for a trajectory ending at $\tau=x+i y$  fixes the parameter $t_0$ in terms of $y$.

The discussion for $M_{c,d}$ is similar. These geometries are obtained from $M_{0,1}$ by acting with an $\r{SL}(2,\Z)$ element corresponding to taking a linear combination of the two circles to be the circle that shrinks smoothly. As a result the set of geometries considered here corresponds to all possible ways of filling the torus. 

By repeating the above analysis for the trajectory \eqref{trajGEN}, we find that  there are cross terms between the two circles. To cancel them we can apply an $\r{SL}(2,\Z)$ transformation $(v,u)\ra (v,u).\g$ with $\g =\big(\begin{smallmatrix} a&b\\ c& d\end{smallmatrix}\big)\in \r{SL}(2,\Z)$. The cross terms can be canceled only if the endpoint $x_0$ is rational of the form $x_0 = -{d\/c}$. Then the no-boundary condition sets
\be
t_0  = {y\/4((cx+d)^2 + c^2 y^2)}~,
\ee
which is as expected the $\r{SL}(2,\Z)$ transform of \eqref{t0M01} under $\g$. The lesson here is that the no-boundary condition gives a strong restriction on the no-boundary trajectories, that they must end up at rational points on the real line. This is equivalent to saying that the no-boundary trajectories must be $\r{SL}(2,\Z)$-equivalent to a trajectory coming from the cusp at $i\infty$, since the rational points are its images under $\r{SL}(2,\Z)$. This interpretation will lead to an interesting representation  in Artin's billiard.

Finally let us discuss the trajectories in flat space coordinates. With the no-boundary condition, the leading trajectory intersects the light-cone $T=0$ at the ``Euclidean point''
\be
\vec{X}_{0,1} = (X_0,Y_0,Z_0) = (0, \tfrac12 y^2,\tfrac12 y^2)~.
\ee
This gives a simple way to compute the action simply from the distance in $\R^{1,2}$:
\be
S_{0,1} = -M D(T,x,y;\vec{X}_0) = -M T + {M y\/2}+O(T^{-1})~.
\ee
The corresponding Euclidean point for the $M_{c,d}$ trajectory is obtained similarly
\be
\vec{X}_{c,d}= {y^2\/2 ((cx+d)^2+c^2 y^2)} \big({-}2cd,d^2-c^2,d^2+c^2\big)~,
\ee
and we can check that the distance reproduces the expected on-shell action
\be
S_{c,d} = -M D(T,x,y;\vec{X}_{c,d}) = -M T + {M y\/2 ((cx+d)^2+c^2 y^2)}+O(T^{-1})~.
\ee
It's interesting to note that a similar structure was found in the canonical quantization of dS$_2$ JT gravity \cite{Maldacena:2019cbz}. The classical solutions there also correspond to trajectories in an auxiliary (two-dimensional) Minkowski space and the no-boundary condition fixes a position on the light-cone.

\ss{Hartle-Hawking in Artin's billiard}\label{sec:Artin}

We have shown that the classical trajectories of the particle corresponds to cosmological spacetimes. As discussed above, the no-boundary condition restricts the trajectories to be $\r{SL}(2,\Z)$-equivalent to a vertical trajectory, corresponding to a particle coming from the cusp at $i\infty$.

\begin{figure}
\begin{center}
	\begin{tabular}{cc}
		\subf{\includegraphics[width=6cm]{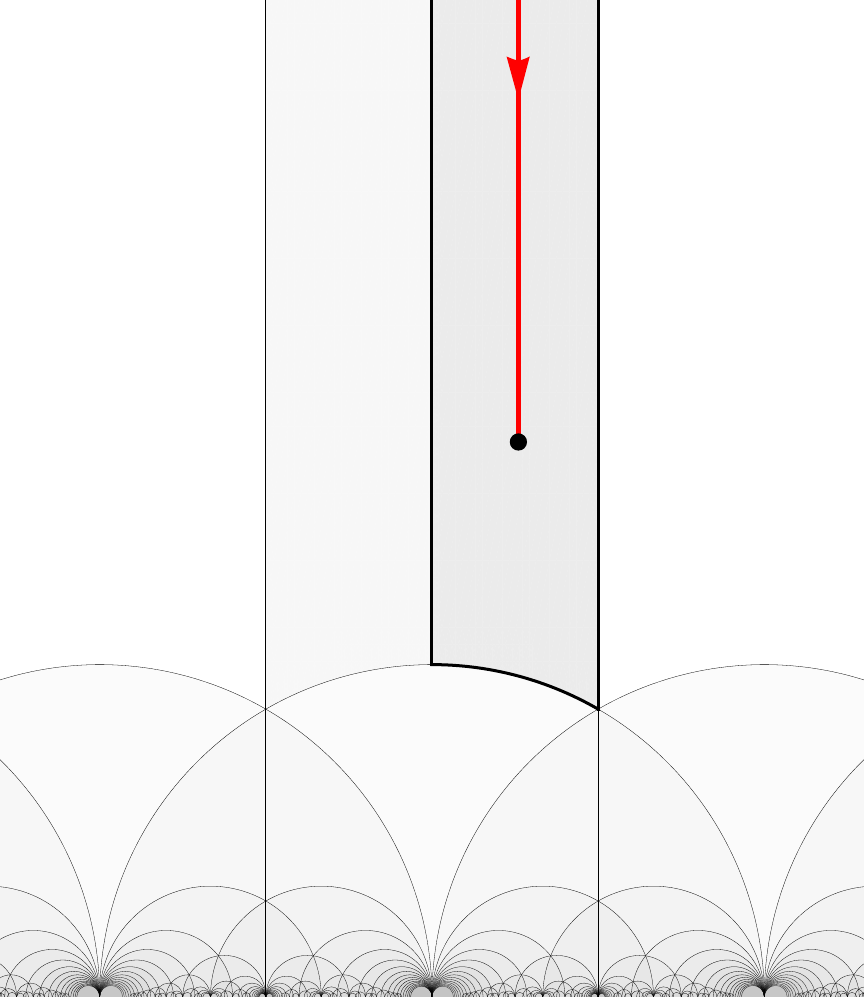}}{a. $M_{0,1}$ }&
		\subf{\includegraphics[width=6cm]{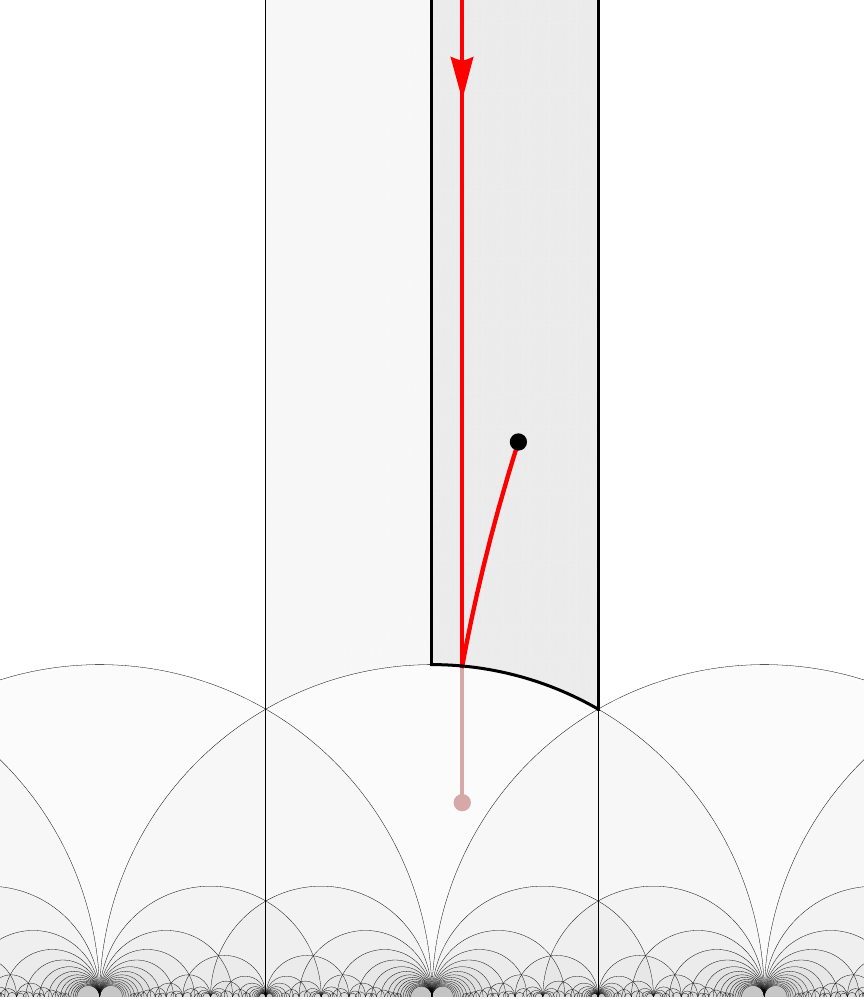}}{b. $M_{1,0}$}  
	\end{tabular}
\end{center}
\caption{Classical spacetimes correspond to particle trajectories in Artin's billiard.}\label{fig:M01M10}
\end{figure}

Now let's discuss the sum over no-boundary trajectories which defines the Hartle-Hawking state. This sum was considered in \cite{Maloney:2007ud,Keller:2014xba} in the context of AdS$_3$ and discussed in \cite{Castro:2012gc} in the cosmological context. The sum over geometries is really a sum over the coset $\G_\infty\backslash\r{PSL}(2,\Z)$. The classical geometries $M_{c,d}$ are in one-to-one correspondence with choices of two coprime integers $(c,d)$ with $c\geq0$. Given two such integers, we can form an $\r{SL}(2,\Z)$ matrix $(\begin{smallmatrix} a & b \\ c & d\end{smallmatrix})$ where the integers $(a,b)$ are uniquely fixed modulo the $\G_\infty$ action generated by shifts $(a,b)\ra (a,b)+(c,d)$. We can always pick a representative satisfying $|{a\/c}|<{1\/2}$ so that the image $\g\tau$ remains in the vertical strip $|x|<{1\/2}$. 

As discussed above, the leading geometry $M_{0,1}$ is just the vertical trajectory going from $i\infty$ to the point $\tau\in\cF$, see Figure~\ref{fig:M01M10}a. To obtain the $M_{c,d}$ trajectory, we act with an element of $\r{SL}(2,\Z)$ on $\tau$. The image $\g\tau$ of $\tau\in \cF$ is not in the fundamental domain  anymore. The trajectory $M_{c,d}$ is the vertical trajectory going from $i\infty$ to $\g\tau$. Although it looks vertical in the covering space $\bH$, it is actually more complicated in $\cF$ since its crosses the boundaries. For example, $M_{1,0}$ includes a bounce on the unit circle corresponding to the application of the $S$ transformation, see Figure~\ref{fig:M01M10}b.

\begin{figure}
\begin{center}
	\begin{tabular}{ccc}
		\subf{\includegraphics[width=4cm]{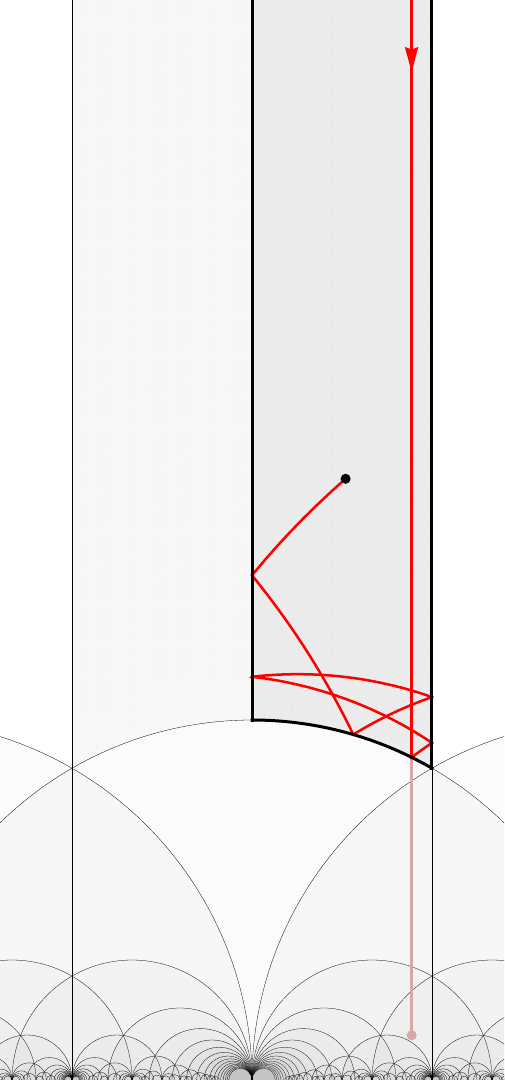}}{$(c,d)=(2,1)$ } &
		\subf{\includegraphics[width=4cm]{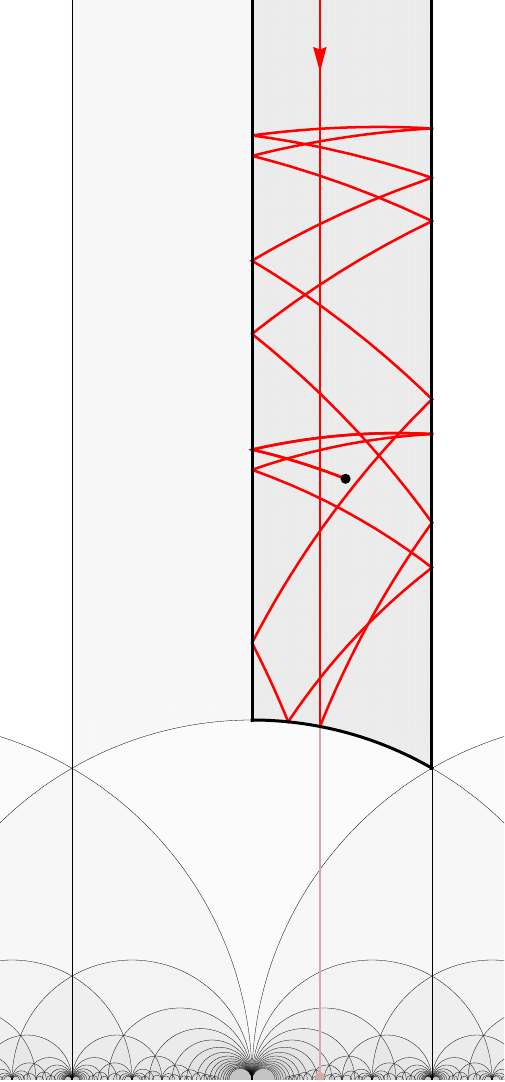}}{$(c,d)=(5,11)$} &
		\subf{\includegraphics[width=4cm]{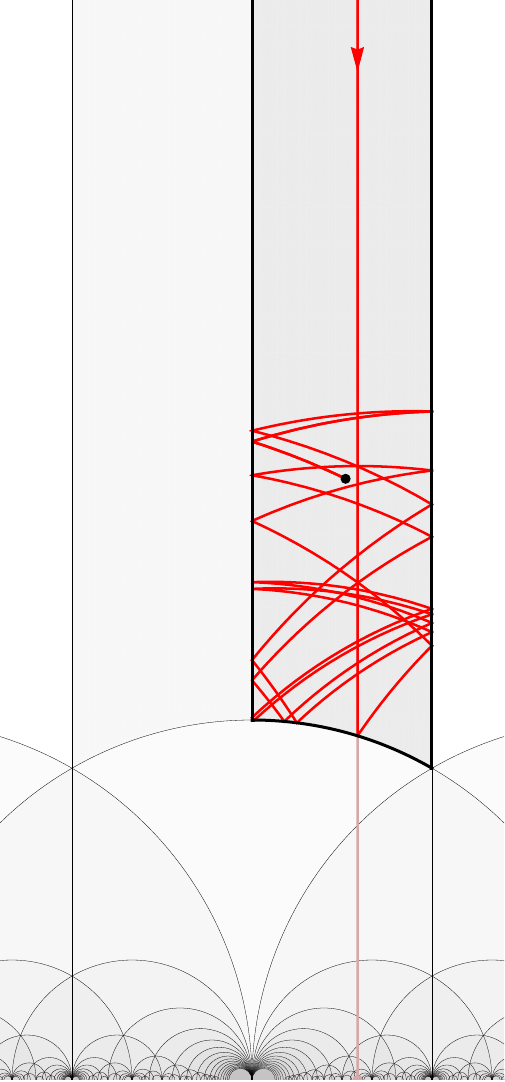}}{$(c,d)=(17,41)$}  
	\end{tabular}
\end{center}
\caption{Some examples of the trajectories  $M_{c,d}$. The Hartle-Hawking state is defined by summing over all these trajectories ending at the black point $\tau\in \cF$ .}\label{fig:Mcd}

\end{figure}

When a trajectory crosses a boundary at $x=\pm {1\/2}$, it comes back to the other side as these two sides are identified. To represent the resulting trajectory, it is convenient to perform a $\Z_2$ quotient corresponding to the reflection  $x\ra-x$. The geodesic then becomes a billiard trajectory in the positive half of the fundamental domain. This is known as Artin's billiard \cite{artinMechanischesSystemMit1924}. Note that this $\Z_2$ identification is natural in gravity since the torus diffeomorphism $(u,v)\ra(u,-v)$  does flip the sign of $x$ and is part of the gauge group.  A general $M_{c,d}$ then corresponds to a trajectory in Artin's billiard as illustrated in Figure~\ref{fig:Mcd}.

It's interesting to ask whether we can predict some of the features of the trajectory from the integers $(c,d)$. In fact, the dynamics can be discretized and corresponds to  iterating the Gauss  map starting from the rational ${c\/d}$. This is an incarnation of the well-studied relationship between $\r{SL}(2,\Z)$, continued fractions, and geodesics on the modular surface, see for example \cite{artinMechanischesSystemMit1924,merrimanCarolineSeriesModular,grabinerCuttingSequencesGeodesic2001,katokSymbolicDynamicsModular2006,katokArithmeticCodingGeodesics,morier-genoudFareyBoatContinued2019}. 

In our context, the main point is that the trajectory corresponds to a decomposition of the $\r{SL}(2,\Z)$ matrix as 
\be
\g = \big( \begin{smallmatrix} a&b\\ c&d \end{smallmatrix} \big) = ST^{n_m} \dots ST^{n_2}S T^{n_1}
\ee
where $S=\big(\begin{smallmatrix} 0 &-1\\ 1& 0 \end{smallmatrix} \big)$ and $T=\big(\begin{smallmatrix} 1&1\\ 0&1 \end{smallmatrix} \big)$ are the standard generators and $(n_1,\dots,n_m)$ are integers encoding the trajectory. This decomposition tells us the list of transformations we have to perform to put back the particle in the fundamental domain after it crosses one of its boundaries. Equivalently, it gives the sequences of images of the fundamental domain  crossed by the vertical trajectory in the covering space.

From the fact that $\g^{-1}(i\infty) = -{d\/c}$, we see that this decomposition leads to the continued fraction representation 
\be
{c\/d}= n_1 - {1\/n_2-{1\/\ddots -{1\/n_{m}}}}~.
\ee
Note that such continued fraction representation or decomposition of $\g$ in $S$ and $T$ are highly non-unique, due to the relations satisfied by the generators. To obtain the integers $n_k$ corresponding to our problem, we can simply follow the trajectory so that it remains in Artin's billiard.

\begin{figure}
	\centering
	\includegraphics[width=12cm]{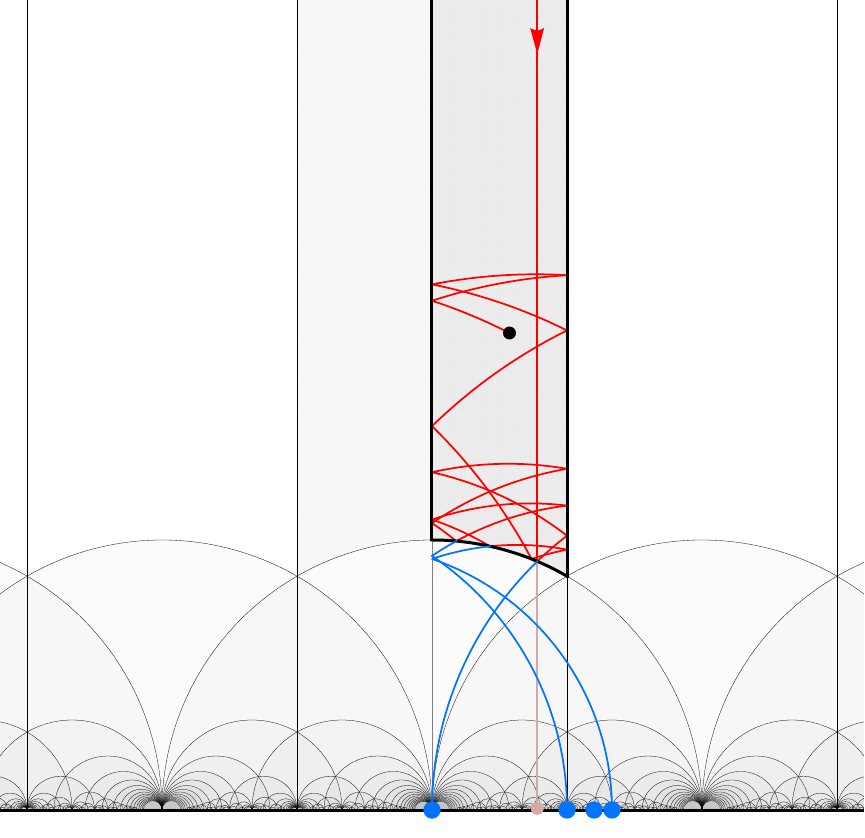}
		\put(-121,-6){${1\/2}$}
		\put(-110,-6){${3\/5}$}
		\put(-102,-6){${2\/3}$}
		\put(-174,-6){$0$}

	\caption{The trajectory for $(c,d)=(5,13)$. There are four bounces on the unit circle whose intersections with the real axis are given by iterating the Gauss map starting from ${c\/d}$.}\label{fig:Gauss}
\end{figure}
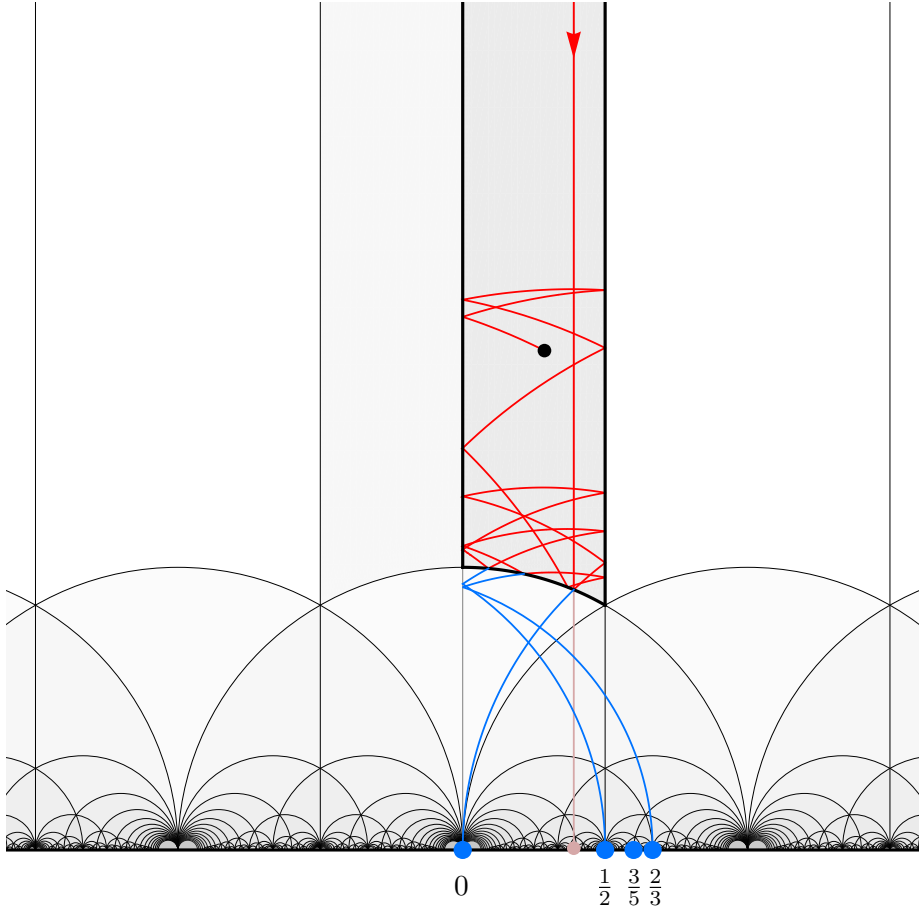

Given a choice of $(c,d)$, the integers $n_k$ can be obtained by iterating the Gauss map defined for $x\in \R$ as
\be
G(x) = {1\/x} - \Big\lfloor {1\/x} \Big\rfloor
\ee
and $G(0)=0$. Here $\lfloor\cdot\rfloor$ is the integer part or floor function. If we consider the sequence of iterates of the Gauss map starting from ${c\/d}$
\be
q_0 = \Big|{c\/d}\Big|,\qq q_{k+1}=G(q_k)~,
\ee
we get a finite sequence of rational numbers. The integers $n_k$ are then obtained as the integer parts of their reciprocals:
\be
n_k= \r{sign}(\tfrac{c}{d})\,  (-1)^{k+1} \Big\lfloor {1\/q_{k-1}}\Big\rfloor~.
\ee  
Note that this algorithm gives the correct trajectory for many choices of $(c,d)$. When ${d\/c}$ is almost an integer, a small modification is necessary at the end of the sequence. The condition is that when $n_{m-1}=-1$, we must use the relation $ST^{-1}S = TST$ to get the new sequence $(n_1,n_2,\dots,n_{m-2}+1,n_m+1)$ of length $m-1$. Ultimately these codings depend on the choice of fundamental domain, and  the Farey tessellation of $\bH$ is more appropriate to give the precise relation in general \cite{merrimanCarolineSeriesModular, morier-genoudFareyBoatContinued2019}. See also \cite{pohlSymbolicDynamicsGeodesic2013,pohlDynamicsGeodesicsMaass2020} for a different coding appropriate for the relation between closed geodesics and Maass cusp forms.

To illustrate this procedure, let us explain how this goes for the choice $(c,d)=(5,13)$ illustrated in Figure~\ref{fig:Gauss}.  By iterating the Gauss map, we get the sequence 
\be
G^{(k)}(\tfrac{5}{13})\quad  = \quad \tfrac{5}{13},\quad  \tfrac{3}{5},\quad \tfrac{2}{3},\quad \tfrac{1}{2},\quad 0,\quad 0,\quad \dots 
\ee
and the above formula gives the integers
\be
(n_1,n_2,n_3,n_4) = (2,-1,1,-2)~,
\ee
corresponding to the decomposition of $\g$ and the continued fraction representation
\be
\g=\bpm 2 & 5 \\ 5 & 13 \epm = S T^{-2} S T ST^{-1} S T^2,\qq {5\/13} = {1\/2-{1\/-1-{1\/1-{1\/-2}}}}~.
\ee
The Gauss map will always give zero after a finite number of iterations when we start with a rational number. The number of non-zero values gives the number of bounces on the unit circle, or equivalently the number of times $S$ appears in the decomposition of $\g$ corresponding to the trajectory. Moreover, the sequence $(q_k)$ gives the positions at which these bounces intersect the real axis. As a result  the Gauss map can be viewed as a discretization of the dynamics. Note that the Gauss map is one of the simplest example of a dynamical system which is chaotic, ergodic and exponentially mixing \cite{CORLESS1990241,exponentialMixing,barreiraErgodicTheoryHyperbolic2012}.

The Hartle-Hawking state is defined by summing over all trajectories coming from the cusp at $i\infty$. As a result, it is the wavefunction of a particle with the boundary condition that it came from $i\infty$. As such, it can be viewed as a notion of quantum vacuum for the Artin particle.

Note that the Hartle-Hawking state is quite special as it only involves rational trajectories, \ie those that are $\r{SL}(2,\Z)$-equivalent to a vertical trajectory on $\bH$.  So this corresponds to geodesics on $\bH$ for which at least one endpoint is rational (or $i\infty$). 
A generic trajectory is irrational and those trajectories are much more complicated in Artin's billiard. One way to see this is that the Gauss map iterates won't become stationary so we expect an infinite number of bounces on the unit circle. In fact, Artin used his billiard to give the first example of a geodesic on a Riemann surface with a dense trajectory \cite{artinMechanischesSystemMit1924}. Here the restriction to rational trajectories is due to the no-boundary condition, \ie that the spacetime has a smooth continuation in Euclidean signature. Irrational trajectories correspond to Lorentzian spacetimes starting at a singularity which cannot be smoothen out by going to Euclidean signature. It would be interesting to see if the dynamics of irrational trajectories in Artin's billiard could help understand the Lorentzian singularity, see \cite{Cornish:1996hx,Damour:2002et,Forte:2008jr,DeClerck:2023fax} for ideas in this direction.

\section{Second quantization}\label{sec:second}

The second-quantized theory is the spacetime theory corresponding to the particle. Thus it is a scalar field of mass $M$ on the auxiliary spacetime
\be\label{auxMetricSecond}
ds^2_\r{Aux}= G_{ab} dX^a dX^b =  dT^2-T^2{dx^2+dy^2\/y^2}~,
\ee
and is described by the action
\bea
\cS \=- {1\/2}\int d^3 X\sqrt{\r{det}\,G}\le(- G^{ab}\p_a\phi\p_b\phi  + M^2\phi^2 \ri)~.
\eea
The equation of motion is the  Klein-Gordon  equation 
\be
(\Box+M^2)\phi=0~.
\ee
The metric \eqref{auxMetricSecond} is flat and corresponds to the hyperbolic slicing of the future diamond of Minkowski space as given in \eqref{XYZtoTxy}. The quantization of a scalar field in this metric was described in \cite{deBoer:2003vf}. Although this is just a free scalar in flat space, the theory is not as simple as it may seem because the particle does not live on $\bH$ but on its $\r{PSL}(2,\Z)$ quotient. Indeed the auxiliary spacetime  is  similar to the manifold $\r{SL}(2,\R)/\r{SL}(2,\Z)$ which is topologically the complement of the trefoil knot \cite{Ghys}.

% Here we have in addition the $\r{PSL}(2,\Z)$ quotient.

The second-quantized theory should really be thought as a third-quantized theory from the perspective of gravity as  discussed for example in \cite{Giddings:1988cx,Giddings:1988wv,Marolf:2020xie}. This formalism is useful because it is more powerful than the first-quantized approach. For example a complicated sum over particle trajectories becomes a two-point function in the second-quantized theory.

\ss{Mode decomposition and harmonic analysis}

 The Klein-Gordon equation takes the form
\be
(D_T + T^2M^2+ \D)\phi=0
\ee
where we have defined $D_T\chi  = \p_T(T^2\p_T\chi)$ and $\D=-y^2(\p_x^2+\p_y^2)$.  The inner product that we should use is the Klein-Gordon inner product
\be\label{KGinner}
(\phi_1,\phi_2) =i \int {dx dy\/y^2}(\phi_1 T^2\p_T\phi_2^\ast- \phi_2 T^2\p_T\phi_1^\ast)~,
\ee
as this is the inner product that is invariant with respect to time evolution, thus ensuring unitarity.  The field operator decomposes as
\be
\phi(T,x,y) = \int_\R d\mu \,(a_\mu^\dg \phi_\mu^++ a_\mu \phi_\mu^-)+\sum_{n\geq 1} (b_n^\dg \varphi_n^++b_n \varphi_n^-)
\ee
and contains  both a continuous and a discrete piece. These modes are orthonormal 
\be
(\phi^+_\mu,\phi^+_\nu) =  \d(\mu-\nu),\qq (\phi^-_\mu,\phi^-_\nu) = -\d(\mu-\nu),\qq (\phi^+_\mu,\phi^-_\nu)=0~,
\ee
and similarly for $\vphi_n^\pm$. In particular we can check the canonical commutation relations. The conjugate momentum is
\be
\pi = {\d S\/\d\p_T \phi} = {T^2\/y^2}\p_T\phi
\ee
and we see that
\be
[\pi(T,x,y),\phi(T,x',y')] =-i \d^{{(2)}}(\tau-\tau')
\ee
implies that we have
\be
[a_\mu^\dg,a_\nu] = \d(\mu-\nu)~,\qq [b_n^\dg,b_m] = \d_{m,n}~.
\ee
As a result, $a_\nu^\dg$ is a creation operator for the wavefunction $\phi_\nu^-$ which describes an expanding universe. The wavefunction $\phi_\nu^+$ is a contracting universe and corresponds to the annihilation operator $a_\nu$. We can define a second-quantized vacuum $|0\rn$ from the condition that $a_\nu|0\rn=b_n|0\rn=0$ for all $\nu$ and $n$.

The elementary modes decompose in a time part and a space part according to
\be
\phi_\nu^\pm(T,x,y)=\chi^\pm_\nu(T) \psi_\nu(x,y)~,
\ee
which satisfy
\be\label{modesEquations}
\D  \psi_\nu(x,y) = \big(\tfrac14+\nu^2\big) \psi_\nu(x,y),\qq
 (D_T+T^2M^2) \chi_\nu^\pm(T)= -\big( \tfrac14+\nu^2\big)\chi_\nu^\pm(T)~.
\ee
The Klein-Gordon inner product \eqref{KGinner} separates into an inner product on time and an inner product on space. If we write
\be
\phi_i(T,x,y) = \chi_i(T) \psi_i(x,y),\qq i=1,2,
\ee
we obtain
\be
(\phi_1,\phi_2) =  (\chi_1,\chi_2)_T \ln \psi_1,\psi_2\rn~.
\ee
The spatial inner product is the Petersson inner product 
\be
\ln\psi_1,\psi_2\rn \equiv \int_\cF{ dx dy\/y^2} \psi_1(x,y)\psi_2^\ast(x,y)~,
\ee
while in time we get 
\bea
(\chi_1,\chi_2)_T\equiv  i (\chi_1(T)T^2\p_T\chi_2^\ast(T)- \chi_2^\ast(T)T^2\p_T\chi_1(T))~.
\eea
The time-dependent modes $\chi_\nu^\pm(T)$ will be discussed in the next section. 

The spatial modes are automorphic  eigenfunctions of the Laplacian. Thus the spatial mode decomposition follows from the well-studied subject of harmonic analysis in the fundamental domain \cite{terrasHarmonicAnalysisSymmetric1985,iwaniecSpectralMethodsAutomorphic2002}. Any element  $\psi\in L^2(\cF)$ can be decomposed as
\be
\psi(x,y) = {1\/4\pi}\int_\R d\nu \,\ln \psi,E_{{1\/2}+i\nu}\rn  E_{{1\/2}+i\nu}(x,y) + \sum_n  \ln \psi,f_n\rn f_n(x,y)~,
\ee
and this basis is orthonormal for the Petersson inner product.  Thus we see that the spatial modes $ \psi_\nu(x,y)$ consist in  Eisenstein series and Maass cusp forms. 

Note that this spectral decomposition has been discussed recently in the context of asymptotically AdS$_3$ quantum gravity \cite{Benjamin:2021ygh,Haehl:2023tkr,Haehl:2023xys}. It was used in particular to study the factorization puzzle \cite{DiUbaldo:2023qli,Haehl:2023mhf}. Since AdS$_3$ partition functions are not in $L^2(\cF)$, a subtraction is necessary to apply the spectral decomposition. In quantum cosmology, the states are defined to be in $L^2(\cF)$ so the spectral decomposition is unavoidable.

Let us review some facts about this  decomposition.  The continuous spectrum is given by the Eisenstein series $E_s(x,y)$ for $s={1\/2}+i\nu,\nu\in\R$ which satisfy
\be
\D E_{{1\/2}+i\nu} = (\tfrac14+\nu^2) E_{{1\/2}+i\nu}~.
\ee
The Eisenstein series can be defined as Poincaré sums
\be
E_s(x,y) = \sum_{\g\in \G_\infty\backslash\G} \r{Im}(\g\tau)^s
\ee
which can be analytically continued to any $s\in \C$. They have the Fourier expansion  \cite{kubotaElementaryTheoryEisenstein1973,ZagierEisenstein,sarnakSpectralTheoryAutomorphic}
\be\label{EisensteinFourier}
E_s(x,y) = y^{s}+\vphi(s)y^{1-s}+{4 \sqrt{y}\/\L(s)}\sum_{n\geq 1}  c_{s-{1\/2}}(n) K_{s-{1\/2}}(2\pi n  y)\,\r{cos}(2\pi n x)
\ee
where the $n$-th Fourier coefficient involves
\be
c_{s}(n) = n^s \sum_{\substack{d\geq 1\\d|n}} d^{-2s}  = \sum_{\substack{ a,d\geq 1\\ad = n}} \Big({a\/d}\Big)^s
\ee
and  the completed Riemann zeta function 
\be
\L(s)=\pi^{-s}\G(s)\z(2s)
\ee 
which satisfies the functional equation $\L(s) = \L(\tfrac12-s)$.  The constant Fourier mode has a reflected piece in terms of the phase 
\be
\vphi(s) = {\L(1-s)\/\L(s)}~,
\ee
and it will be useful to note that 
\be
\vphi(\tfrac12+i\nu) = {\L(\tfrac12-i\nu)\/\L(\tfrac12+i\nu)} ={\L(i\nu)\/\L(-i\nu)}~.
\ee
The Eisenstein series satisfy the functional equation
\be
\L(s) E_s(x,y) = \L(1-s)E_{1-s}(x,y)~.
\ee
Note that the Eisenstein series are not strictly in $L^2(\cF)$. Their inner product is divergent but can be regularized by cutting the fundamental domain to some height and using the Maass-Selberg relations \cite{kubotaElementaryTheoryEisenstein1973,iwaniecSpectralMethodsAutomorphic2002,sarnakSpectralTheoryAutomorphic}.

The discrete spectrum $\{f_n\}$ consists of truly $L^2$ functions, known as Maass cusp forms. They are special and rather mysterious eigenfunctions of the Laplacian \cite{muhlenbruchTheoryMaassWave2020} which can only be studied  indirectly using the Selberg trace formula \cite{Selberg,hejhalSelbergTraceFormula1976} or the Selberg zeta function \cite{zbMATH04205458,Ruelle,FriedZeta}. The discrete spectrum is conjectured to be simple \cite{sarnakSpectraHyperbolicSurfaces2003} and the corresponding eigenvalues are only known numerically \cite{Hejhal1991}. These functions are of interest in number theory because they are related to $L$-functions. Here we see that  they appear naturally as quantum states of pure three-dimensional cosmology.

Maass cusp forms are pseudo-random objects \cite{luoQuantumErgodicityEigenfunctions1995,2005math6102F} so we might expect the corresponding eigenvalues to follow random matrix statistics. This is true for the Laplace spectrum of generic Riemann surfaces but it is false here. Indeed the case at hand is special because the modular surface $\bH/\r{PSL}(2,\Z)$ is arithmetic and its spectrum follows Poissonian statistics. This is due to the existence of Hecke operators which provide an integrability-like structure \cite{PhysRevLett.69.2188}.  Hecke operators act as unitaries on the Hilbert space so it would be interesting to see if they can be given a physical interpretation in the present context and if one can understand the integrability structure they realize. See also \cite{laxScatteringTheoryAutomorphic1980,Gutzwiller_Jose_1990} for  relevant ideas about automorphic dynamics.

\ss{Time evolution and Hamiltonian}

The elementary solutions of \eqref{modesEquations} can be written in terms of Hankel functions with imaginary order
\be
\chi_\mu^+(T)= e^{-{\pi\mu\/2}}\sqrt{\pi \/4 T} H^{(1)}_{i\mu}(MT) ,\qq \chi_\mu^-(T)=e^{{\pi\mu\/2}} \sqrt{\pi \/4 T} H^{(2)}_{i\mu}(MT)~.
\ee
We note that they are complex conjugates of each other and symmetric
\be
(\chi_\mu^+)^\ast = \chi_\mu^-\qq (\chi_\mu^-)^\ast =  \chi_\mu^+,\qq \chi_{-\mu}^+ = \chi_\mu^+ ,\qq \chi_{-\mu}^- =\chi_\mu^-~,
\ee
and have a normalized inner product
\be
(\chi_\mu^+,\chi_\mu^+)_T = 1,\qq (\chi_\mu^-,\chi_\mu^-)_T = -1,\qq (\chi_\mu^+,\chi_\mu^-)_T=0~.
\ee
Let's discuss the asymptotic behavior of these modes. At late time we have
\be
\chi_\mu^\pm (T) \sim {1\/\sqrt{2M} T} e^{\pm i (MT-{\pi\/4})},\qq T\to+\infty~.
\ee
In particular we note that $\chi_\mu^\pm(T)$ captures the counterterms in the wavefunctional discussed above. From the sign of the divergent phase, we see that $\chi_\mu^-(T)$ corresponds to an expanding universe while $\chi_\mu^+(T)$ is a contracting one.

In the limit $T\to0$ we have an oscillatory behavior
\be\label{chiearly}
\chi_\mu^\pm(T)\sim \mp{i\/2}\sqrt{M\/2\pi}\le(  e^{\pm {\pi\mu\/2}} \G(-i\mu) (\tfrac12 MT)^{-{1\/2}+i\mu} +e^{\mp{\pi\mu\/2}} \G(i\mu) (\tfrac12 MT)^{-{1\/2}-i\mu}\ri),\quad T\to0~.
\ee
This is expected since $T\to0$ corresponds to the singularity.

We can also consider the second-quantized Hamiltonian which takes the form
\be
\cH_T ={1\/2}\int {dx dy\/y^2}\le( {y^4\/T^2}\pi^2 + \phi \Delta\phi + T^2 M^2\phi^2 \ri)~.
\ee
Note that it depends on the slice $T=\r{const}$ on which we evaluate it. Hence it is a time-dependent Hamiltonian. From the mode decomposition, we obtain
\be
\cH_T = {1\/2}\int d\mu \le( \w_T(\mu) a_\mu^\dg a_\mu +\hc\ri)~,
\ee
in terms of the energy density
\be
\w_T(\mu ) =T^2 \p_T\chi_\mu^+(T) \p_T\chi_\mu^-(T)+ \le( \tfrac14+\mu^2 + T^2 M^2\ri) \chi_\mu^+(T)\chi_\mu^-(T)~.
\ee
At late and early times, this takes the form
\bea
\w_{T\to\infty}(\mu)  \= M  + O(T^{-2})~,\\
\w_{T\to 0}(\mu)  \= {(\tfrac14+\mu^2)\/\mu\,\r{tanh}(\pi \mu) T} + O(1) + \text{oscillatory}~.
\eea
At early times, the energy density is peaked around $\mu=0$ but becomes uniform under time evolution. Thus the system has a  diffusive/dissipative behavior.

The second-quantized formulation suggests a way to define  a kind of cosmological S-matrix. In terms of the Hamiltonian it can be written as the unitary evolving from $T=0$ to $T=\infty$:
\be
S = \cT \,\r{exp}\le( i\int_0^{+\infty} dT \,\cH_T\ri)~.
\ee
 In the Minkowski embedding, this is an S-matrix where the initial state is defined on the future light-cone and the final state is at future infinity.

\ss{Green function and automorphic kernels}

A sum over particle trajectories computes the Green function in the second-quantized formalism. The auxiliary spacetime is flat so the Green function is especially simple. The standard Green function for a massive scalar field in $\R^{1,2}$ gives
\be
\cG_0(\vec{X},\vec{X}_0) ={1\/D}\,e^{-i  MD}
\ee
where $D$ is the distance
\be
D(\vec{X},\vec{X}_0) = \sqrt{(Z-Z_0)^2-(X-X_0)^2-(Y-Y_0)^2}~.
\ee
which after writing in $(T,x,y)$ coordinates gives at late time
\be
\cG_0(T,x,y;T_0,x_0,y_0)=  {1\/T}\,\r{exp}\le( -i MT+ i MT_0{(x-x_0)^2 +y^2\/ 2 y y_0} \ri),\qq T\to+\infty~.
\ee
The divergent terms are the counterterms discussed above so the finite Green function can be defined as
\be\label{GreenK}
G_0(x,y;T_0,x_0,y_0)=\lim_{T\to\infty}T e^{iMT} \cG_0(T,x,y;T_0,x_0,y_0)=  e^{i MT_0\,\r{cosh}\,d(\tau,\tau_0)}~,
\ee
where $d(\tau,\tau_0)$ is the hyperbolic distance on $\bH$:
\be
\r{cosh}\,d(\tau,\tau_0) = 1+\tfrac12 t(\tau,\tau_0),\qq t(\tau,\tau_0)={|\tau-\tau_0|^2\/ y y_0}~.
\ee
In particular, this reproduces the $\TT$ kernel given in \eqref{TTkernel}. This is expected as the Green function of the second-quantized theory captures time evolution of the particle theory.

The above discussion was for the covering space  so we still have to sum over $\r{PSL}(2,\Z)$ images. The Green function is then
\be
G(\tau;T_0,\tau_0) = \sum_{\g\in \G} G_0(\tau;T_0,\g\tau_0)~.
\ee
More generally, for any function $k(t)$ of the point-pair invariant $t(\tau,\tau_0)$, we can consider the automorphic kernel
\be
K(\tau,\tau_0)=\sum_{\g\in \G} k(t(\tau,\g\tau_0))~.
\ee
These automorphic kernels  are the operators acting on the first-quantized Hilbert space.

Quantities of this type can be studied using the spectral decomposition \cite{Selberg}. For any point-pair invariant $k(t)$, the kernel can be written as
\be
K(\tau,\tau_0) = {1\/4\pi}\int_\R d\nu\,h(\nu) E_{{1\/2}+i\nu}(\tau)E_{{1\/2}-i\nu}(\tau_0)~,
\ee
where the spectral density $h(\nu)$ is obtained from $k(t)$ by the so-called Selberg/Harish-Chandra transform \cite{kubotaElementaryTheoryEisenstein1973,iwaniecSpectralMethodsAutomorphic2002}. Let us briefly recall this procedure. From $k(t)$, we first define a function $Q(w)$ as
\be
\int_w^{+\infty} dt\,{k(t)\/\sqrt{t-w}} = Q(w)~,
\ee
and then the function
\be\label{defg}
g(u) = Q(2\,\r{cosh}\,u-2)~.
\ee
The spectral density $h(\nu)$ is then obtained by Fourier transform
\be
h(\nu) = \int_\R du\,g(u)\,e^{i\nu u}~.
\ee
For the Green function \eqref{GreenK} we have
\be
k(t)=e^{i M T_0} e^{{iM T_0\/2} t}~,
\ee
so we can compute $Q(w)$ and we get
\be
g(u) = e^{i\pi\/4}\sqrt{2\pi\/M T_0}\, e^{i M T_0\,\r{cosh}\,u}~.
\ee
The Fourier transform with respect to $u$ then shows that the spectral density is given by a Hankel function
\be\label{GreenHankel}
h(\nu) = \int du\,e^{i u\nu} g(u) = i\pi e^{-\pi\nu/2}e^{i\pi\/4}\sqrt{2\pi\/M T_0} H_{i\nu}^{(1)}(MT_0)~.
\ee
It is instructive to match this with a direct computation using the scalar field operator. The quantity of interest is simply the two-point function 
\be
\cG(T,x,y;T_0,x_0,y_0) = \ln 0| \phi(T,x,y)\phi(T_0,x_0,y_0)|0\rn~.
\ee
Using the mode decomposition, we obtain 
\be
\cG(T,x,y;T_0,x_0,y_0) = \int_\R d\nu\, \chi^-_\nu(T) \chi^+_\nu(T_0)  E_{{1\/2}+i\nu}(\tau) E_{{1\/2}-i\nu}(\tau_0)
\ee
so this gives the spectral decomposition of the Green function. In the limit $T\to\infty$, what remains is
\be
\lim_{T\to+\infty} T e^{i MT}\chi_\nu^-(T)\chi_\nu^+(T_0) = {1\/\sqrt{2M} } \, e^{-{i\pi\/4}} e^{-{\pi\nu\/2}}\sqrt{\pi \/4 T_0} H^{(1)}_{i\nu}(MT_0) = {1\/4\pi} h(\nu)
\ee
which precisely reproduces the expected density.

We are particularly interested in the limit $T_0\to0$ corresponding to an initial condition for universes of zero size. We can define states by integrating the Green function over a source $\la_{T_0}(x_0,y_0)$ and taking the limit $T_0\to 0$. In this case the Hankel function behaves as \eqref{chiearly}. So we need a source with a particular dependence on $T_0$ as to get a finite limit $T_0\to0$. This is the second-quantized version of the no-boundary condition discussed previously in the first-quantized setting.

\ss{The Hartle-Hawking spectrum}

The Hartle-Hawking state is defined as 
\be
\Psi_\r{HH}(\tau)=\sum_{\g\in \G_\infty\backslash\G} \Psi_{0,1}(\g\tau),\qq\Psi_{0,1}(\tau)= \sqrt{y}|q|^{-{c-1\/12}}|1-q|^2~,
\ee
using $q=e^{2i\pi\tau}$ with $\tau=x+iy$. This is the same as the Maloney-Witten partition function \cite{Maloney:2007ud}. As explained above we have multiplied by $\sqrt{y}|\eta(\tau)|^2$ to focus on the primary partition function which corresponds to removing the Virasoro modes. This combination is modular invariant and doesn't affect the Poincaré sum.

In section \ref{sec:Artin}, we have seen that the Hartle-Hawking state can be viewed as the sum over rational particle trajectories, \ie trajectories coming from the cusp at $i\infty$. As such the Hartle-Hawking state appears to define a notion of vacuum for the particle theory. However, this picture is insufficient because it only captures the leading contribution $|q|^{-{c-1\/12}}$ and the one-loop piece $\sqrt{y}|1-q|^2$ is crucial to obtain a well-defined Poincaré sum. This one-loop piece reflects the fact that the CFT vacuum lies in a degenerate representation due to $\r{SL}(2,\R)$ zero modes. In the second-quantized theory, the effect of the one-loop piece can be interpreted as a non-trivial source that can be derived from the spectral decomposition.

The Hartle-Hawking spectral density is defined as the Rankin-Selberg transform of $\Psi_\r{HH}$, \ie as the Petersson inner product with the Eisenstein series:
\be
\rho_\r{HH}(\nu) = (\Psi_\r{HH},E_{{1\/2}+i\nu}) = \int_\cF {dxdy\/y^2} \Psi_\r{HH}(x,y) E_{{1\/2}-i\nu}(x,y)~.
\ee
Since the Hartle-Hawking state is expressed as a Poincaré sum, we can unfold the integration domain from the fundamental domain $\cF$ to the strip $|x|<{1\/2}$. This gives the formula
\be\label{rhoHHdef}
\rho_\r{HH}(\nu) =\int_0^{+\infty }{dy\/y^2} \int_{-{1\/2}}^{{1\/2}}dx\,\Psi_{0,1}(x,y) E_{{1\/2}-i\nu}(x,y)~.
\ee
To perform this integral, we will parametrize the central charge as
\be
c= 1-6 Q^2 = 25-6 \wt{Q}^2,\qq Q  = b-{1\/b},\qq\wt{Q} = b+{1\/b}~.
\ee
This parametrization is purely for convenience. We expect that the Hartle-Hawking state can be defined for any complex value of $c$. The Poincaré sum definition is manifestly convergent in the region $\r{Re}(c)<1$ which corresponds to $b$ real so we can define the wavefunction in this  region and analytically continue it in $c$. The AdS case corresponds to $c={3\/2G}+O(1)$ real and large. As discussed above, in the cosmological context we need
\be
c={3i\/2G}+13 + O(1)i
\ee
where we allow only purely imaginary corrections. This corresponds to taking
\be
b = e^{\pm {i\pi\/4}}\la,\qq\la\in \R~.
\ee
We expect that our expressions will be valid for generic values of the central charge if we define them by analytic continuation from convergent regions where $\r{Re}(c)$ is sufficiently small. Hence our considerations will also apply to the AdS$_3$ context.

Explicitly the seed can be written as a sum of four exponentials
\be
\Psi_{0,1} = \sqrt{y}( e^{- Q^2 \pi y}+ e^{-\wt{Q}^2 \pi y}-2 e^{-(b^2+{1\/b^2})\pi y} \,\r{cos}(2\pi x))~,
\ee
and only the constant and first Fourier coefficients of the Eisenstein series contribute 
\bea
E_{{1\/2}-i\nu}(x,y) \= y^{{1\/2}-i\nu}+{\L(-i\nu)\/\L(i\nu)}y^{{1\/2}+i\nu}+{4\sqrt{y}\/\L(i\nu)} K_{-i\nu}(2\pi  y) \,\r{cos}(2\pi  x)+\dots
\eea
The contribution from the constant Fourier modes follows from the Cahen–Mellin integral
\be
\int_0^{+\infty} dy\,e^{-\a y}y^{s-1} = \a^{-s} \G(s)~.
\ee
For the first Fourier mode, we  can use the integral representation 
\be
K_\nu(z) = \tfrac12 (\tfrac12 z)^\nu \int_0^{+\infty} {dt\/t^{1+\nu}} \,\r{exp}\Big({-}t -{z^2\/4t}\Big)~,
\ee
to obtain the identity
\bea\label{Kidentity}
{1\/\pi}\int_\R d\mu \,\D^{i\mu} K_{i\mu}(2\pi y ) \= {1\/2\pi}\int_0^{+\infty} {dt\/t}\int_\R d\mu \,\Big( {\pi y \D \/t}\Big)^{i\mu}  \,\r{exp}\Big({-}t -{\pi^2 y^2\/t}\Big)~, \-
\= \r{exp}\Big({-}\pi y\Big(\D+{1\/\D}\Big)\Big)~.
\eea
The Kontorovich–Lebedev transform then gives the answer for the integral that we want to evaluate:
\be
\int_0^{+\infty} {dy\/y}\, K_{i\mu}(2\pi y)  \,\r{exp}\Big({-}\pi y\Big(\D+{1\/\D}\Big)\Big) = {\pi\/2\mu \,\r{sinh}(\pi\mu)} (\D^{i\mu}+\D^{-i\mu})~.
\ee
Note that the application of the Kontorovich–Lebedev  transform is formal here since the LHS has a pole at $y=0$ which should be removed. The RHS should be viewed as the regularized definition. As we will see, this turns out to be equivalent to the regularization used by Maloney-Witten.

Our final result for the Hartle-Hawking spectral density takes the form
\be\label{HHdensity}
\rho_\r{HH}(\nu)  = \pi^{i\nu}\G(-i\nu) \le[ Q^{2i\nu}+\wt{Q}^{2i\nu} + {\z(-2i\nu)\/\z(2i\nu)}(Q^{-2i\nu}+\wt{Q}^{-2i\nu} ) - {2\/\z(2i\nu) }(b^{2i\nu}+b^{-2i\nu}) \ri]
\ee
so that the spectral representation of the Hartle-Hawking state is
\be\label{HHspectraldef}
\Psi_\r{HH}(x,y) = {1\/4\pi}\int_\R d\nu\,\rho_\r{HH}(\nu) E_{{1\/2}+i\nu}(x,y)~.
\ee
Note that, as any Rankin-Selberg transform, $\rho_\r{HH}$ satisfies the functional equation
\be
\rho_\r{HH}(\nu) = \vphi(\tfrac12+i\nu)\rho_\r{HH}(-\nu)~,
\ee
which follows from the corresponding equation for the Eisenstein series. It appears more appropriate to think of $\rho_\r{HH}$ as an $L$-function rather than as a density. In particular, it would be interesting to see if $\rho_\r{HH}$ admits an Euler product formula.

This result also gives the source necessary to produce the Hartle-Hawking state following the discussion in the previous section. The effect of time evolution corresponds to the Hankel function \eqref{GreenHankel} which produces the factor $\G(-i\nu)$ in the $T_0\to0$ limit. So up to scaling by a $T_0$-dependent prefactor necessary to get a finite $T_0\to0$ limit, the Hartle-Hawking source is essentially the expression in bracket in \eqref{HHdensity}.

As a consistency check, we can compare to the expansion of the constant Fourier mode obtained by  Maloney-Witten  \cite{Maloney:2007ud}:
\be
\Psi_\r{HH}^{(0)} = \sum_{m\geq 0} w_m \,y^{{1\/2}-m}~,
\ee
where  $w_0=-6$ and
\be
w_m = {\pi^{m+{1\/2}}\/ m\G(m+\tfrac12)\z(2m+1)}\le[ \z(2m)\Big((\tfrac{c-1}{6})^m + (\tfrac{c-25}{6})^m\Big) - 4T_m(\tfrac{c-13}{12})\ri]~,
\ee
with $T_m$ the Chebyshev polynomial of the first kind.

From our spectral representation, we see that the constant Fourier mode is
\be
\Psi_\r{HH}^{(0)} = {1\/4\pi}\int_\R d\nu \,\rho_\r{HH}(\nu) (y^{{1\/2}+i\nu}+\vphi(\tfrac12+i\nu) y^{{1\/2}-i\nu}) = {1\/2\pi} \int_\R d\nu\,\rho_\r{HH}(\nu) \,y^{{1\/2}+i\nu}~,
\ee
where we used the functional equation to rewrite it as a single term. Now we can close the contour by picking up all the poles with $\r{Im}(\nu)>0$. The gamma function only has poles for $\r{Im}(\nu)\leq 0$ so the only poles in \eqref{HHdensity} come from the trivial zeros of the zeta function $\nu = i n , n\geq 1$. The non-trivial zeros of the zeta function also don't contribute as the critical strip corresponds to $-{1\/2}<\r{Im}(\nu) <0$.

This gives the expansion
\be
\Psi_\r{HH}^{(0)} = \sum_{n\geq 1} \,\r{Res}_{\nu = i n}\,\rho_\r{HH}(\nu)\,y^{{1\/2}-n}
\ee
and we can check that the residues precisely match the Maloney-Witten coefficients
\be
\r{Res}_{\nu = i n} \,\rho_\r{HH}(\nu)=  w_n~.
\ee
In particular, the Chebyshev polynomial is correctly reproduced using the identity
\be
T_m(\tfrac12(x+x^{-1})) = {1\/2}(x^m+x^{-m})~,
\ee
as we have ${c-13\/12} = -{1\/2}(b^2+b^{-2})$ in our parametrization.

Interestingly the constant term $w_0=-6$ is also correctly reproduced from the pole of the gamma function at $\nu=0$. For this note that given the presence of a pole at $\nu=0$, the contour of integration should be shifted to  $\nu\in \R+i\e$. Then when combining $\rho_\r{HH}$ with its reflection, only one of the two contributes after we close the contour so the constant contribution is
\be
{1\/2}\times \r{Res}_{\nu=0}\,  \rho_\r{HH}(\nu)= -6~.
\ee
Thus the density \eqref{HHdensity} is precisely what is needed to reproduce the Maloney-Witten expansion of the constant Fourier mode. This implies that the spectral representation \eqref{HHspectraldef} matches the Maloney-Witten partition function as the constant Fourier mode uniquely fixes the Eisenstein spectrum. In principle, discrete contributions due to Maass cusp forms could be present but studying this is more difficult and cannot be done with the present considerations, as Maass cusp forms have a vanishing constant Fourier mode.

\ss{Hartle-Hawking as a Möbius average}

In the AdS context, the meaning of the Maloney-Witten partition function remains a subject of debate due to its unphysical features, notably an apparently continuous spectrum and non-positive density of states. See  \cite{Castro:2011zq,
Keller:2014xba,Benjamin:2019stq,Maxfield:2020ale,Alday:2019vdr,Kaidi:2020ecu,
Benjamin:2020mfz,Meruliya:2021utr,Mertens:2022ujr,Chandra:2022bqq,DiUbaldo:2023hkc} for a sample of discussions on the subject. In the cosmological context, it is viewed as a  wavefunction so there is a priori no requirement for either positivity or discreteness.

Interestingly the spectral representation derived in the previous section gives a way to rewrite the Maloney-Witten partition function as a $q$-expansion with integer coefficients. This is surprising but is possible here because we allow negative degeneracies and accumulation points in the spectrum. As we will explain, this representation gives a precise way to view the Hartle-Hawking state, or the Maloney-Witten partition function in the AdS context, as an average.

We can separate the Hartle-Hawking state in three terms, each separately modular invariant:
\be
\Psi_\r{HH} = \psi[Q]+\psi[\wt{Q}] -\chi[b]~,
\ee
corresponding to different pieces of \eqref{HHdensity}
\bea
\psi[Q] \= {1\/4\pi}\int_\R d\nu \,\pi^{i\nu}\G(-i\nu) \le[ Q^{2i\nu}+{\z(-2i\nu)\/\z(2i\nu)}Q^{-2i\nu}\ri] E_{{1\/2}+i\nu}~,\-
\chi[b] \= {1\/2\pi}\int_\R d\nu \,\pi^{i\nu}\G(-i\nu) \le[  {1\/\z(2i\nu) }(b^{2i\nu}+b^{-2i\nu})\ri] E_{{1\/2}+i\nu}~.
\eea
They separately compute the Poincaré sums appearing in the Hartle-Hawking state:
\bea
\psi[Q] \= \sum_{\g\in \G_\infty\backslash\G} \sqrt{y} |q|^{Q^2\/2}\big|_\g ,\qq \chi[b] = 2\sum_{\g\in \G_\infty\backslash\G} \sqrt{y} |q|^{ {1\/2}(b^2+{1\/b^2})}\,\r{cos}(2\pi x)\big|_\g~,
\eea
where $q=e^{2i\pi\tau}$ and $|_\g$ denotes the action of $\g$ on $\tau=x+iy$.

The main idea is that  we can rewrite them as discrete sums of exponentials by using the Dirichlet series representations of the zeta function
\be
\z(-2i\nu) = \sum_{n\geq 1} n^{2i\nu},\qq {1\/\z(2i\nu)} = \sum_{m\geq 1}\mu(m) m^{-2i\nu}~.
\ee
The series for the reciprocal of the zeta function involves the Möbius function $\mu(m)$ defined as $\mu(m)=(-1)^k$ when $m$ is the
product of $k$ distinct primes and 0 when $m$ contains a square factor. 

For example the constant Fourier coefficient of $\psi[Q]$ gives
\bea
\psi^{(0)}[Q] \= {1\/2\pi}\int_\R d\nu\,\pi^{i\nu} \G(-i\nu)\Big( Q^{2i\nu} +\sum_{m,n\geq 1}\mu(m)  \Big({n\/Qm}\Big)^{2i\nu} \Big) y^{{1\/2}+i\nu}~,\-
\= \sqrt{y} |q|^{Q^2\/2} + \sqrt{y}  \sum_{m,n\geq 1}\mu(m) |q|^{{n^2\/2m^2 Q^2}}~,
\eea
which becomes a discrete sum of exponentials. The non-zero Fourier coefficients can also be computed from the Fourier decomposition \eqref{EisensteinFourier} of the Eisenstein series and using repeatedly the identity \eqref{Kidentity}.

After the dust settles we obtain the expression
\be
Z_\r{HH} ={1\/\sqrt{y}|\eta(\tau)|^2}\le( \psi[Q]+\psi[\wt{Q}]-\chi[b]\ri)
\ee
where we write here the full partition function including the Virasoro prefactor. The different contributions take the form
\bea
\psi[Q] \= \sqrt{y}\,|q|^{Q^2/2}+ \sqrt{y} \sum_{m,n \geq 1}  \mu(m) |q|^{{n^2\/2m^2 Q^2}} \-
&& \hspace{3cm}+\sqrt{y}\sum_{m\geq 1}\mu(m)\sum_{a,d\geq 1}( q^{h_{m,a,d}^+ }\bar{q}^{h^-_{m,a,d}}+\bar{q}^{h_{m,a,d}^+ }q^{h^-_{m,a,d}})~,\\
\chi[b] \=  \sqrt{y} \sum_{m\geq 1}  \mu(m)(|q|^{ {b^2\/2m^2} }+|q|^{ {1\/2m^2b^2} })  \-
&&\hspace{3cm} +\sqrt{y} \sum_{m,\l\geq 1}\mu(m)\mu(\l) \sum_{a,d\geq 1}( q^{h^+_{m,\l,a,d}}\bar{q}^{h^-_{m,\l,a,d}}+\bar{q}^{h^+_{m,\l,a,d}}q^{h^-_{m,\l,a,d}})\Big]~,
\eea
and the conformal dimensions that appear are
\begin{align}\label{confdim}
h^\pm_{m,a,d}  = {1\/4} (p^\pm_{m,a,d})^2, \qq \wt{h}^\pm_{m,a,d}  = {1\/4} (\wt{p}^\pm_{m,a,d})^2 \qq h^\pm_{m,\l,a,d} & = {1\/4} (p^\pm_{m,\l, a,d})^2,
\end{align}
which we have parametrized in terms of  Liouville momenta
\be\label{liouvillemomenta}
p^\pm_{m,a,d}  = a m Q\pm d {1\/m Q},\qq \wt{p}^\pm_{m,a,d}  = a m \wt{Q}\pm d {1\/m \wt{Q}},\qq p^\pm_{m,\l,a,d} = a {\l b\/m}  \pm d {m\/\l b}~.
\ee
The expansion of $\psi[\wt{Q}]$ is the same as that of $\psi[Q]$ but with $Q$ replaced by $\wt{Q}$. For $\chi[b]$, the sum over $m,\l$ has an apparent divergence at the diagonal $m=\l$ but this should be zeta-regularized using $\sum_{m\geq 1}|\mu(m)| m^{-s} ={\z(s)\/\z(2s)} $ so that $\sum_{m\geq 1}\mu(m)^2=1$.

This gives an exact  expression for the Maloney-Witten partition function as a $q$-expansion with integer coefficients. The existence of such an expression  is surprising but is possible here as it involves a rather exotic spectrum with negative degeneracies and accumulation points.

The signs of the degeneracies are governed by the Möbius function which is valued in $\{-1,0,1\}$. This function is central in analytic number theory and is known to display chaotic behavior \cite{iwaniec2004analytic,GreenTao,Kaisa}. In particular, its sign  is essentially random as it is believed that
\be\label{mobsum}
\sum_{n\leq N}\mu(n) = O(N^{{1\/2}+\e})~,
\ee
which is what would be expected from a random walk. Such pseudo-random behavior is intimately related to the distribution of prime numbers and in fact \eqref{mobsum} is equivalent to the Riemann hypothesis. Additional conjectures by Chowla and Sarnak suggest that the Möbius function is as random as one could expect \cite{TaoBlogPost}.  The ``Möbius randomness hypothesis'' is the idea  that summing $\mu(n)$ against any reasonable function will lead to significant cancellations \cite{sarnakThreeLecturesMobius}.

From this perspective the Hartle-Hawking state can be viewed as a random average obtained as follows. For the term $\psi[Q]$ and $\psi[\wt{Q}]$, we replace $Q\ra m Q$ and $\wt{Q}\ra m \wt{Q}$. For the term $\chi[b]$, we replace  $b\ra {\l\/m} b$. We then perform the average by summing over $m,\l$ with the Möbius function. This can be viewed as an average over the central charge. From the pseudo-random nature of the Möbius function we expect large cancellations ensuring convergence. It would be interesting to make this more precise and identify the seed CFT that is averaged here. The conformal dimensions \eqref{confdim} suggest that degenerate Virasoro representations make an appearance but there are also other operators in the spectrum.

\section{Future directions}

This paper has only explored the simplest case of pure three-dimensional quantum cosmology with a spatial Riemann surface with $g=1$. A similar story should exist for $\cM_{g,n}$ corresponding to a surface of genus $g$ with $n$ perturbative insertions. Adding operator insertions (metric or matter fluctuations) gives a way to define a perturbative Hilbert space, as described in \cite{Chakraborty:2023yed}. In three dimensions it should be the Hilbert space of Virasoro blocks discussed for example in \cite{Eberhardt:2023mrq}. It would be interesting to develop this more general story. The case of the sphere $g=0$ presents additional challenges due to the existence of a cosmological horizon and a static patch.

The equivalence with the particle suggests  an avenue to understand off-shell gravity in three dimensions \cite{Eberhardt:2022wlc,Maxfield:2020ale}. As  classical spacetimes correspond to classical particle trajectories, a consistent prescription to sum over off-shell  spacetimes would be  to sum over all particle trajectories. Such particle path integrals do make sense and can be studied more easily  in the second-quantized formalism. 

In particular it is interesting to think about closed geodesics on moduli space which would correspond to time-periodic cosmologies. They cannot arise as on-shell spacetimes but they could appear as off-shell contributions in the second-quantized theory. The Selberg zeta function relates Maass cusp forms to the lengths of closed geodesics on the modular surface \cite{Lewis1997PeriodFA,ZagierLewisMaaass,ZagierSelbergZeta,pohlDynamicsGeodesicsMaass2020} so it would be interesting to obtain gravitational incarnations of this and other zeta functions \cite{hejhalSelbergTraceFormula1976,urgenFischer,eiseleDynamicalZetaFunctions1994,Ruelle}.

In this paper, we have focused on the semi-classical story. Non-perturbative effects include for example topology change. In the particle picture, this would correspond to a particle jumping from one topological sector to another. Whether such effects can be made sense of is an interesting avenue for future research. The gravity path integral gives a way to study this by integrating over suitable cobordisms. In particular, it would be interesting to give an interpretation in our context of the wormholes discussed in \cite{Maldacena:2004rf,Cotler:2020ugk,Yan:2023rjh}.

Recently it has been suggested that the Hilbert space of quantum cosmology is one-dimensional so that the theory would be trivial. In particular this follows from arguments based on factorization \cite{Penington:2019kki,McNamara:2020uza,Benini:2022hzx,Usatyuk:2024isz}. The Hilbert space discussed in this paper is infinite-dimensional as well as the one constructed in \cite{Chakraborty:2023yed}. There is no discrepancy because a Hilbert space depends on a choice of inner product and such choice must be viewed as part of the definition of the theory. The inner product appearing naturally in considerations about factorization may not be the one we should use to discuss cosmology.  This point was already made by Higuchi in the perturbative de Sitter case \cite{Higuchi:1991tk} where a redefinition of the inner product was necessary to obtain a non-trivial Hilbert space \cite{Higuchi:1991tm}. This redefinition is natural because it corresponds to dividing by the infinite volume of the residual conformal group   \cite{Chakraborty:2023los} and necessary in order to have an interesting theory of quantum cosmology. 

An important but conceptually more difficult question concerns the role of the observer. The perspective taken in this paper is that of the meta-observer who is probing the universe externally. Instead, we have arisen as emergent observers in the universe and we are probing it from the inside. Discussing quantum mechanics in this context is more challenging, see \cite{Witten:2023qsv,Witten:2023xze} for some recent ideas about this.

Finally we would like to comment on the potential connections to number theory. Modular forms are familiar to physicists as they arise as partition functions of two-dimensional CFTs. However, these are mostly not the modular forms that are of interest in number theory as they behave badly at the cusps and hence don't give rise to $L$-functions. Instead, the automorphic forms appearing as quantum states in cosmology are precisely the ones studied by number theorists because by definition we want the wavefunctions to be in $L^2$.

Our real interest is to understand quantum cosmology in higher dimensions. The ideas and tools used in this paper could be applied more generally.  In higher dimensions, wavefunctionals in each topological sector should also be automorphic forms but of more general types. The existence of the $\TT$ deformation in higher dimensions \cite{Taylor:2018xcy,Hartman:2018tkw} and its expected relation to the Laplace operator suggest that, at least in the large volume regime, the dynamics will also involve harmonic analysis on moduli space. These are the ingredients appearing in the Langlands program \cite{jacquet1970automorphic,Bump_1997}. It would be interesting to understand  whether Hecke operators or Galois representations can be realized in quantum cosmology and if semi-classical gravity can be related to number-theoretic trace formulas. From a physics perspective,  number theory is interesting because it provides a wealth of pseudo-random objects, some of which having already appeared in this work. Despite displaying chaotic behavior, these quantities are also exact, so it is natural to speculate that they could play a role in a microscopic description of gravity.

\appendix

\acknowledgments

I would like to thank the  ICTS string theory community and especially Suvrat Raju for many stimulating discussions  on topics related to quantum cosmology. This work also benefited from discussions with P. Lebard, Kevin Nguyen and Gabriele Di Ubaldo. I am also grateful for the hospitality of KITP Santa Barbara, where part of this work was done during the workshop ``What is String Theory?''.
 I acknowledge support from the ERC-COG grant NP-QFT No. 864583
“Non-perturbative dynamics of quantum fields: from new deconfined
phases of matter to quantum black holes”, by the MUR-FARE2020 grant
No. R20E8NR3HX “The Emergence of Quantum Gravity from Strong Coupling
Dynamics” and by INFN Iniziativa Specifica GAST.

\bibliographystyle{JHEP}
\bibliography{refs}

\providecommand{\href}[2]{#2}\begingroup\raggedright\begin{thebibliography}{100}

\bibitem{Witten:2001kn}
E.~Witten, {\it {Quantum gravity in de Sitter space}},  in {\em {Strings 2001:
  International Conference}}, 6, 2001.
\newblock \href{http://arxiv.org/abs/hep-th/0106109}{{\tt hep-th/0106109}}.

\bibitem{DeWitt:1967yk}
B.~S. DeWitt, {\it {Quantum Theory of Gravity. 1. The Canonical Theory}},  {\em
  Phys. Rev.} {\bf 160} (1967) 1113--1148.

\bibitem{Deser:1973zza}
S.~Deser and D.~Brill, {\it {Instability of Closed Spaces in General
  Relativity}},  {\em Commun. Math. Phys.} {\bf 32} (1973) 291.

\bibitem{Moncrief:1976un}
V.~Moncrief, {\it {Space-Time Symmetries and Linearization Stability of the
  Einstein Equations. 2.}},  {\em J. Math. Phys.} {\bf 17} (1976) 1893--1902.

\bibitem{Higuchi:1991tk}
A.~Higuchi, {\it {Quantum linearization instabilities of de Sitter space-time.
  1}},  {\em Class. Quant. Grav.} {\bf 8} (1991) 1961--1981.

\bibitem{Higuchi:1991tm}
A.~Higuchi, {\it {Quantum linearization instabilities of de Sitter space-time.
  2}},  {\em Class. Quant. Grav.} {\bf 8} (1991) 1983--2004.

\bibitem{Arnowitt:1962hi}
R.~L. Arnowitt, S.~Deser, and C.~W. Misner, {\it {The Dynamics of general
  relativity}},  {\em Gen. Rel. Grav.} {\bf 40} (2008) 1997--2027,
  [\href{http://arxiv.org/abs/gr-qc/0405109}{{\tt gr-qc/0405109}}].

\bibitem{Chakraborty:2023yed}
T.~Chakraborty, J.~Chakravarty, V.~Godet, P.~Paul, and S.~Raju, {\it {The
  Hilbert space of de Sitter quantum gravity}},  {\em JHEP} {\bf 01} (2024)
  132, [\href{http://arxiv.org/abs/2303.16315}{{\tt arXiv:2303.16315}}].

\bibitem{Chakraborty:2023los}
T.~Chakraborty, J.~Chakravarty, V.~Godet, P.~Paul, and S.~Raju, {\it
  {Holography of information in de Sitter space}},  {\em JHEP} {\bf 12} (2023)
  120, [\href{http://arxiv.org/abs/2303.16316}{{\tt arXiv:2303.16316}}].

\bibitem{Hartle:1983ai}
J.~B. Hartle and S.~W. Hawking, {\it {Wave Function of the Universe}},  {\em
  Phys. Rev. D} {\bf 28} (1983) 2960--2975.

\bibitem{Strominger:2001pn}
A.~Strominger, {\it {The dS / CFT correspondence}},  {\em JHEP} {\bf 10} (2001)
  034, [\href{http://arxiv.org/abs/hep-th/0106113}{{\tt hep-th/0106113}}].

\bibitem{Maldacena:2002vr}
J.~M. Maldacena, {\it {Non-Gaussian features of primordial fluctuations in
  single field inflationary models}},  {\em JHEP} {\bf 05} (2003) 013,
  [\href{http://arxiv.org/abs/astro-ph/0210603}{{\tt astro-ph/0210603}}].

\bibitem{Maldacena:2019cbz}
J.~Maldacena, G.~J. Turiaci, and Z.~Yang, {\it {Two dimensional Nearly de
  Sitter gravity}},  {\em JHEP} {\bf 01} (2021) 139,
  [\href{http://arxiv.org/abs/1904.01911}{{\tt arXiv:1904.01911}}].

\bibitem{Maldacena:2024uhs}
J.~Maldacena, {\it {Comments on the no boundary wavefunction and slow roll
  inflation}},  \href{http://arxiv.org/abs/2403.10510}{{\tt arXiv:2403.10510}}.

\bibitem{deserThreedimensionalCosmologicalGravity1984}
S.~Deser and R.~Jackiw, {\it Three-dimensional cosmological gravity: dynamics
  of constant curvature},  {\em Annals of Physics} {\bf 153} (1984), no.~2
  405–416.

\bibitem{Witten:1988hc}
E.~Witten, {\it {(2+1)-Dimensional Gravity as an Exactly Soluble System}},
  {\em Nucl. Phys. B} {\bf 311} (1988) 46.

\bibitem{Witten:1989ip}
E.~Witten, {\it {Quantization of {Chern-Simons} Gauge Theory With Complex Gauge
  Group}},  {\em Commun. Math. Phys.} {\bf 137} (1991) 29--66.

\bibitem{verlindeConformalFieldTheory1989}
E.~Verlinde and H.~Verlinde, {\it Conformal field theory and geometric
  quantization},  vol.~Proc. Superstrings, p.~422–449, World Scientific,
  1989.

\bibitem{Maldacena:1998ih}
J.~M. Maldacena and A.~Strominger, {\it {Statistical entropy of de Sitter
  space}},  {\em JHEP} {\bf 02} (1998) 014,
  [\href{http://arxiv.org/abs/gr-qc/9801096}{{\tt gr-qc/9801096}}].

\bibitem{Witten:2007kt}
E.~Witten, {\it {Three-Dimensional Gravity Revisited}},
  \href{http://arxiv.org/abs/0706.3359}{{\tt arXiv:0706.3359}}.

\bibitem{martinecSolubleSystemsQuantum1984}
E.~Martinec, {\it Soluble systems in quantum gravity},  {\em Physical Review D}
  {\bf 30} (Sept., 1984) 1198–1204.

\bibitem{hosoyaDimensionalPureGravity1990}
A.~Hosoya and K.-i. Nakao, {\it (2+ 1)-dimensional pure gravity for an
  arbitrary closed initial surface},  {\em Classical and Quantum Gravity} {\bf
  7} (1990), no.~2 163.

\bibitem{carlipNotesDimensionalWheelerDeWitt1994}
S.~Carlip, {\it Notes on the (2+1)-dimensional wheeler-dewitt equation},  {\em
  Classical and Quantum Gravity} {\bf 11} (Jan., 1994) 31–39.

\bibitem{Moncrief}
V.~Moncrief, {\it {Reduction of the Einstein equations in 2+ 1 dimensions to a
  Hamiltonian system over Teichmüller space}},  {\em Journal of mathematical
  physics} {\bf 30} (1989), no.~12 2907–2914.

\bibitem{seriuBackReactionTopological1996}
M.~Seriu, {\it Back reaction on the topological degrees of freedom in
  (2+1)-dimensional spacetime},  {\em Physical Review D} {\bf 53} (Feb., 1996)
  1889–1906.

\bibitem{seriuPartitionFunctionDimensional1997}
M.~Seriu, {\it Partition function for (2+1)-dimensional einstein gravity},
  {\em Physical Review D} {\bf 55} (Jan., 1997) 781–790.

\bibitem{Carlip:2004ba}
S.~Carlip, {\it {Quantum gravity in 2+1 dimensions: The Case of a closed
  universe}},  {\em Living Rev. Rel.} {\bf 8} (2005) 1,
  [\href{http://arxiv.org/abs/gr-qc/0409039}{{\tt gr-qc/0409039}}].

\bibitem{Araujo-Regado:2022jpj}
G.~Araujo-Regado, {\it {Holographic Cosmology on Closed Slices in 2+1
  Dimensions}},  \href{http://arxiv.org/abs/2212.03219}{{\tt
  arXiv:2212.03219}}.

\bibitem{Hikida:2022ltr}
Y.~Hikida, T.~Nishioka, T.~Takayanagi, and Y.~Taki, {\it {CFT duals of
  three-dimensional de Sitter gravity}},  {\em JHEP} {\bf 05} (2022) 129,
  [\href{http://arxiv.org/abs/2203.02852}{{\tt arXiv:2203.02852}}].

\bibitem{Chen:2022ozy}
H.-Y. Chen and Y.~Hikida, {\it {Three-Dimensional de Sitter Holography and Bulk
  Correlators at Late Time}},  {\em Phys. Rev. Lett.} {\bf 129} (2022), no.~6
  061601, [\href{http://arxiv.org/abs/2204.04871}{{\tt arXiv:2204.04871}}].

\bibitem{Chen:2022xse}
H.-Y. Chen, S.~Chen, and Y.~Hikida, {\it {Late-time correlation functions in
  dS$_{3}$/CFT$_{2}$ correspondence}},  {\em JHEP} {\bf 02} (2023) 038,
  [\href{http://arxiv.org/abs/2210.01415}{{\tt arXiv:2210.01415}}].

\bibitem{Chen:2024vpa}
H.-Y. Chen, Y.~Hikida, Y.~Taki, and T.~Uetoko, {\it {Semi-classical saddles of
  three-dimensional gravity via holography}},
  \href{http://arxiv.org/abs/2403.02108}{{\tt arXiv:2403.02108}}.

\bibitem{York:1972sj}
J.~W. York, Jr., {\it {Role of conformal three geometry in the dynamics of
  gravitation}},  {\em Phys. Rev. Lett.} {\bf 28} (1972) 1082--1085.

\bibitem{Zamolodchikov:2004ce}
A.~B. Zamolodchikov, {\it {Expectation value of composite field T anti-T in
  two-dimensional quantum field theory}},
  \href{http://arxiv.org/abs/hep-th/0401146}{{\tt hep-th/0401146}}.

\bibitem{Smirnov:2016lqw}
F.~A. Smirnov and A.~B. Zamolodchikov, {\it {On space of integrable quantum
  field theories}},  {\em Nucl. Phys. B} {\bf 915} (2017) 363--383,
  [\href{http://arxiv.org/abs/1608.05499}{{\tt arXiv:1608.05499}}].

\bibitem{Mazenc:2019cfg}
E.~A. Mazenc, V.~Shyam, and R.~M. Soni, {\it {A $T \bar{T}$ Deformation for
  Curved Spacetimes from 3d Gravity}},
  \href{http://arxiv.org/abs/1912.09179}{{\tt arXiv:1912.09179}}.

\bibitem{McGough:2016lol}
L.~McGough, M.~Mezei, and H.~Verlinde, {\it {Moving the CFT into the bulk with
  $ T\overline{T} $}},  {\em JHEP} {\bf 04} (2018) 010,
  [\href{http://arxiv.org/abs/1611.03470}{{\tt arXiv:1611.03470}}].

\bibitem{Witten:2022xxp}
E.~Witten, {\it {A note on the canonical formalism for gravity}},  {\em Adv.
  Theor. Math. Phys.} {\bf 27} (2023), no.~1 311--380,
  [\href{http://arxiv.org/abs/2212.08270}{{\tt arXiv:2212.08270}}].

\bibitem{Araujo-Regado:2022gvw}
G.~Araujo-Regado, R.~Khan, and A.~C. Wall, {\it {Cauchy slice holography: a new
  AdS/CFT dictionary}},  {\em JHEP} {\bf 03} (2023) 026,
  [\href{http://arxiv.org/abs/2204.00591}{{\tt arXiv:2204.00591}}].

\bibitem{Chen:2023eic}
D.~Chen, X.~Jiang, and H.~Yang, {\it {Holographic TT\textasciimacron{} deformed
  entanglement entropy in dS3/CFT2}},  {\em Phys. Rev. D} {\bf 109} (2024),
  no.~2 026011, [\href{http://arxiv.org/abs/2307.04673}{{\tt
  arXiv:2307.04673}}].

\bibitem{Castro:2012gc}
A.~Castro and A.~Maloney, {\it {The Wave Function of Quantum de Sitter}},  {\em
  JHEP} {\bf 11} (2012) 096, [\href{http://arxiv.org/abs/1209.5757}{{\tt
  arXiv:1209.5757}}].

\bibitem{Maloney:2007ud}
A.~Maloney and E.~Witten, {\it {Quantum Gravity Partition Functions in Three
  Dimensions}},  {\em JHEP} {\bf 02} (2010) 029,
  [\href{http://arxiv.org/abs/0712.0155}{{\tt arXiv:0712.0155}}].

\bibitem{Aharony:2018bad}
O.~Aharony, S.~Datta, A.~Giveon, Y.~Jiang, and D.~Kutasov, {\it {Modular
  invariance and uniqueness of $T\bar{T}$ deformed CFT}},  {\em JHEP} {\bf 01}
  (2019) 086, [\href{http://arxiv.org/abs/1808.02492}{{\tt arXiv:1808.02492}}].

\bibitem{Cardy:2018sdv}
J.~Cardy, {\it {The $ T\overline{T} $ deformation of quantum field theory as
  random geometry}},  {\em JHEP} {\bf 10} (2018) 186,
  [\href{http://arxiv.org/abs/1801.06895}{{\tt arXiv:1801.06895}}].

\bibitem{Datta:2018thy}
S.~Datta and Y.~Jiang, {\it {$T\bar{T}$ deformed partition functions}},  {\em
  JHEP} {\bf 08} (2018) 106, [\href{http://arxiv.org/abs/1806.07426}{{\tt
  arXiv:1806.07426}}].

\bibitem{Cardy:2022mhn}
J.~Cardy, {\it {$T \overline{T}$-deformed modular forms}},  {\em Commun. Num.
  Theor. Phys.} {\bf 16} (2022), no.~3 435--457,
  [\href{http://arxiv.org/abs/2201.00478}{{\tt arXiv:2201.00478}}].

\bibitem{Freidel:2008sh}
L.~Freidel, {\it {Reconstructing AdS/CFT}},
  \href{http://arxiv.org/abs/0804.0632}{{\tt arXiv:0804.0632}}.

\bibitem{Datta:2021kha}
S.~Datta and Y.~Jiang, {\it {Characters of irrelevant deformations}},  {\em
  JHEP} {\bf 07} (2021) 162, [\href{http://arxiv.org/abs/2104.00281}{{\tt
  arXiv:2104.00281}}].

\bibitem{He:2024pbp}
M.~He, {\it {One-loop partition functions in $T\overline{T}$-deformed
  AdS$_3$}},  \href{http://arxiv.org/abs/2401.09879}{{\tt arXiv:2401.09879}}.

\bibitem{Iliesiu:2020zld}
L.~V. Iliesiu, J.~Kruthoff, G.~J. Turiaci, and H.~Verlinde, {\it {JT gravity at
  finite cutoff}},  {\em SciPost Phys.} {\bf 9} (2020) 023,
  [\href{http://arxiv.org/abs/2004.07242}{{\tt arXiv:2004.07242}}].

\bibitem{Polyakov:1987zb}
A.~M. Polyakov, {\it {Quantum Gravity in Two-Dimensions}},  {\em Mod. Phys.
  Lett. A} {\bf 2} (1987) 893.

\bibitem{Nguyen:2021pdz}
K.~Nguyen, {\it {Holographic boundary actions in AdS$_{3}$/CFT$_{2}$
  revisited}},  {\em JHEP} {\bf 10} (2021) 218,
  [\href{http://arxiv.org/abs/2108.01095}{{\tt arXiv:2108.01095}}].

\bibitem{deBoer:2023lrd}
J.~de~Boer, V.~Godet, J.~Kastikainen, and E.~Keski-Vakkuri, {\it {Quantum
  information geometry of driven CFTs}},  {\em JHEP} {\bf 09} (2023) 087,
  [\href{http://arxiv.org/abs/2306.00099}{{\tt arXiv:2306.00099}}].

\bibitem{quineZetaRegularizedProducts1993}
J.~R. Quine, S.~H. Heydari, and R.~Y. Song, {\it Zeta regularized products},
  {\em Transactions of the American Mathematical Society} {\bf 338} (1993),
  no.~1 213–231.

\bibitem{Giombi:2008vd}
S.~Giombi, A.~Maloney, and X.~Yin, {\it {One-loop Partition Functions of 3D
  Gravity}},  {\em JHEP} {\bf 08} (2008) 007,
  [\href{http://arxiv.org/abs/0804.1773}{{\tt arXiv:0804.1773}}].

\bibitem{Carlip:2005tz}
S.~Carlip, {\it {Dynamics of asymptotic diffeomorphisms in (2+1)-dimensional
  gravity}},  {\em Class. Quant. Grav.} {\bf 22} (2005) 3055--3060,
  [\href{http://arxiv.org/abs/gr-qc/0501033}{{\tt gr-qc/0501033}}].

\bibitem{Witten:1987ty}
E.~Witten, {\it {Coadjoint Orbits of the Virasoro Group}},  {\em Commun. Math.
  Phys.} {\bf 114} (1988) 1.

\bibitem{Alekseev:1988ce}
A.~Alekseev and S.~L. Shatashvili, {\it {Path Integral Quantization of the
  Coadjoint Orbits of the Virasoro Group and 2D Gravity}},  {\em Nucl. Phys. B}
  {\bf 323} (1989) 719--733.

\bibitem{Alekseev:1990mp}
A.~Alekseev and S.~L. Shatashvili, {\it {From geometric quantization to
  conformal field theory}},  {\em Commun. Math. Phys.} {\bf 128} (1990)
  197--212.

\bibitem{Coussaert:1995zp}
O.~Coussaert, M.~Henneaux, and P.~van Driel, {\it {The Asymptotic dynamics of
  three-dimensional Einstein gravity with a negative cosmological constant}},
  {\em Class. Quant. Grav.} {\bf 12} (1995) 2961--2966,
  [\href{http://arxiv.org/abs/gr-qc/9506019}{{\tt gr-qc/9506019}}].

\bibitem{Cotler:2018zff}
J.~Cotler and K.~Jensen, {\it {A theory of reparameterizations for AdS$_3$
  gravity}},  {\em JHEP} {\bf 02} (2019) 079,
  [\href{http://arxiv.org/abs/1808.03263}{{\tt arXiv:1808.03263}}].

\bibitem{Nguyen:2022xsw}
K.~Nguyen, {\it {Virasoro blocks and the reparametrization formalism}},  {\em
  JHEP} {\bf 04} (2023) 143, [\href{http://arxiv.org/abs/2212.02527}{{\tt
  arXiv:2212.02527}}].

\bibitem{Marolf:2008it}
D.~Marolf and I.~Morrison, {\it {Group Averaging of massless scalar fields in
  1+1 de Sitter}},  {\em Class. Quant. Grav.} {\bf 26} (2009) 035001,
  [\href{http://arxiv.org/abs/0808.2174}{{\tt arXiv:0808.2174}}].

\bibitem{Chandrasekaran:2022cip}
V.~Chandrasekaran, R.~Longo, G.~Penington, and E.~Witten, {\it {An algebra of
  observables for de Sitter space}},  {\em JHEP} {\bf 02} (2023) 082,
  [\href{http://arxiv.org/abs/2206.10780}{{\tt arXiv:2206.10780}}].

\bibitem{polchinskiEvaluationOneLoop1986}
J.~Polchinski, {\it Evaluation of the one loop string path integral},  {\em
  Communications in Mathematical Physics} {\bf 104} (Mar., 1986) 37–47.

\bibitem{Keller:2014xba}
C.~A. Keller and A.~Maloney, {\it {Poincare Series, 3D Gravity and CFT
  Spectroscopy}},  {\em JHEP} {\bf 02} (2015) 080,
  [\href{http://arxiv.org/abs/1407.6008}{{\tt arXiv:1407.6008}}].

\bibitem{artinMechanischesSystemMit1924}
E.~Artin, {\it Ein mechanisches system mit quasiergodischen bahnen},  {\em
  Abhandlungen aus dem Mathematischen Seminar der Universität Hamburg} {\bf 3}
  (Dec., 1924) 170–175.

\bibitem{merrimanCarolineSeriesModular}
C.~Series, {\it {The Modular Surface and Continued Fractions}},  {\em Journal
  of the London Mathematical Society} {\bf s2-31} (02, 1985) 69--80.
  \url{https://academic.oup.com/jlms/article-pdf/s2-31/1/69/2655643/s2-31-1-69.pdf}.

\bibitem{grabinerCuttingSequencesGeodesic2001}
D.~J. Grabiner and J.~C. Lagarias, {\it Cutting sequences for geodesic flow on
  the modular surface and continued fractions},  {\em Mh Math 133, 295–339
  (2001)} (Apr., 2001). \url{http://arxiv.org/abs/math/9707215}.

\bibitem{katokSymbolicDynamicsModular2006}
S.~Katok and I.~Ugarcovici, {\it Symbolic dynamics for the modular surface and
  beyond},  {\em Bulletin of the American Mathematical Society} {\bf 44} (Oct.,
  2006) 87–132.

\bibitem{katokArithmeticCodingGeodesics}
S.~Katok and I.~Ugarcovici, {\it Arithmetic coding of geodesics on the modular
  surface via continued fractions},  {\em CWI tracts, 2005}.

\bibitem{morier-genoudFareyBoatContinued2019}
S.~Morier-Genoud and V.~Ovsienko, {\it Farey boat: Continued fractions and
  triangulations, modular group and polygon dissections},  {\em Jahresbericht
  der Deutschen Mathematiker-Vereinigung} {\bf 121} (June, 2019) 91–136.
  \url{http://link.springer.com/10.1365/s13291-019-00197-7}.

\bibitem{pohlSymbolicDynamicsGeodesic2013}
A.~D. Pohl, {\it Symbolic dynamics for the geodesic flow on two-dimensional
  hyperbolic good orbifolds},  {\em Discrete and Continuous Dynamical Systems}
  {\bf 34} (2014), no.~5 2173--2241.
  \url{https://www.aimsciences.org/article/id/11bd0561-2b41-4c17-9b64-57b903f1c832}.

\bibitem{pohlDynamicsGeodesicsMaass2020}
A.~Pohl and D.~Zagier, {\it {Dynamics of geodesics, and Maass cusp forms}},
  \href{http://arxiv.org/abs/1906.01067}{{\tt arXiv:1906.01067}}.

\bibitem{CORLESS1990241}
R.~Corless, G.~Frank, and J.~Graham, {\it Chaos and continued fractions},  {\em
  Physica D: Nonlinear Phenomena} {\bf 46} (1990), no.~2 241--253.

\bibitem{exponentialMixing}
M.~Pollicott, {\em Exponential Mixing: Lectures from Mumbai}, pp.~135--167.
\newblock 01, 2019.

\bibitem{barreiraErgodicTheoryHyperbolic2012}
L.~Barreira, {\em Ergodic Theory, Hyperbolic Dynamics and Dimension Theory}.
\newblock Universitext. Springer Berlin Heidelberg, Berlin, Heidelberg, 2012.
\newblock \url{https://link.springer.com/10.1007/978-3-642-28090-0}.

\bibitem{Cornish:1996hx}
N.~J. Cornish and J.~J. Levin, {\it {The Mixmaster universe: A Chaotic Farey
  tale}},  {\em Phys. Rev. D} {\bf 55} (1997) 7489--7510,
  [\href{http://arxiv.org/abs/gr-qc/9612066}{{\tt gr-qc/9612066}}].

\bibitem{Damour:2002et}
T.~Damour, M.~Henneaux, and H.~Nicolai, {\it {Cosmological billiards}},  {\em
  Class. Quant. Grav.} {\bf 20} (2003) R145--R200,
  [\href{http://arxiv.org/abs/hep-th/0212256}{{\tt hep-th/0212256}}].

\bibitem{Forte:2008jr}
L.~A. Forte, {\it {Arithmetical Chaos and Quantum Cosmology}},  {\em Class.
  Quant. Grav.} {\bf 26} (2009) 045001,
  [\href{http://arxiv.org/abs/0812.4382}{{\tt arXiv:0812.4382}}].

\bibitem{DeClerck:2023fax}
M.~De~Clerck, S.~A. Hartnoll, and J.~E. Santos, {\it {Mixmaster chaos in an AdS
  black hole interior}},  \href{http://arxiv.org/abs/2312.11622}{{\tt
  arXiv:2312.11622}}.

\bibitem{deBoer:2003vf}
J.~de~Boer and S.~N. Solodukhin, {\it {A Holographic reduction of Minkowski
  space-time}},  {\em Nucl. Phys. B} {\bf 665} (2003) 545--593,
  [\href{http://arxiv.org/abs/hep-th/0303006}{{\tt hep-th/0303006}}].

\bibitem{Ghys}
E.~Ghys, {\it Knots and dynamics},  {\em Proceedings of the International
  Congress of Mathematicians, Vol. 1, 2006-01-01, ISBN 978-3-03719-022-7, pags.
  247-277} {\bf 1} (01, 2006).
  \url{https://perso.ens-lyon.fr/ghys/articles/knotsdynamics.pdf}.

\bibitem{Giddings:1988cx}
S.~B. Giddings and A.~Strominger, {\it {Loss of incoherence and determination
  of coupling constants in quantum gravity}},  {\em Nucl. Phys. B} {\bf 307}
  (1988) 854--866.

\bibitem{Giddings:1988wv}
S.~B. Giddings and A.~Strominger, {\it {Baby Universes, Third Quantization and
  the Cosmological Constant}},  {\em Nucl. Phys. B} {\bf 321} (1989) 481--508.

\bibitem{Marolf:2020xie}
D.~Marolf and H.~Maxfield, {\it {Transcending the ensemble: baby universes,
  spacetime wormholes, and the order and disorder of black hole information}},
  {\em JHEP} {\bf 08} (2020) 044, [\href{http://arxiv.org/abs/2002.08950}{{\tt
  arXiv:2002.08950}}].

\bibitem{terrasHarmonicAnalysisSymmetric1985}
A.~Terras, {\em Harmonic Analysis on Symmetric Spaces and Applications I}.
\newblock Springer New York, New York, NY, 1985.

\bibitem{iwaniecSpectralMethodsAutomorphic2002}
H.~Iwaniec, {\em Spectral Methods of Automorphic Forms}, vol.~53 of {\em
  Graduate Studies in Mathematics}.
\newblock American Mathematical Society, Providence, Rhode Island, Nov., 2002.

\bibitem{Benjamin:2021ygh}
N.~Benjamin, S.~Collier, A.~L. Fitzpatrick, A.~Maloney, and E.~Perlmutter, {\it
  {Harmonic analysis of 2d CFT partition functions}},  {\em JHEP} {\bf 09}
  (2021) 174, [\href{http://arxiv.org/abs/2107.10744}{{\tt arXiv:2107.10744}}].

\bibitem{Haehl:2023tkr}
F.~M. Haehl, C.~Marteau, W.~Reeves, and M.~Rozali, {\it {Symmetries and
  spectral statistics in chaotic conformal field theories}},  {\em JHEP} {\bf
  07} (2023) 196, [\href{http://arxiv.org/abs/2302.14482}{{\tt
  arXiv:2302.14482}}].

\bibitem{Haehl:2023xys}
F.~M. Haehl, W.~Reeves, and M.~Rozali, {\it {Symmetries and spectral statistics
  in chaotic conformal field theories. Part II. Maass cusp forms and arithmetic
  chaos}},  {\em JHEP} {\bf 12} (2023) 161,
  [\href{http://arxiv.org/abs/2309.00611}{{\tt arXiv:2309.00611}}].

\bibitem{DiUbaldo:2023qli}
G.~Di~Ubaldo and E.~Perlmutter, {\it {AdS$_{3}$/RMT$_{2}$ duality}},  {\em
  JHEP} {\bf 12} (2023) 179, [\href{http://arxiv.org/abs/2307.03707}{{\tt
  arXiv:2307.03707}}].

\bibitem{Haehl:2023mhf}
F.~M. Haehl, W.~Reeves, and M.~Rozali, {\it {Euclidean wormholes in
  two-dimensional conformal field theories from quantum chaos and number
  theory}},  {\em Phys. Rev. D} {\bf 108} (2023), no.~10 L101902,
  [\href{http://arxiv.org/abs/2309.02533}{{\tt arXiv:2309.02533}}].

\bibitem{kubotaElementaryTheoryEisenstein1973}
T.~Kubota, {\em Elementary theory of Eisenstein series}.
\newblock Kodansha scientific books. Kodansha [u.a.], Tokyo, 1973.

\bibitem{ZagierEisenstein}
D.~Zagier, {\it {Eisenstein Series and the Selberg Trace Formula I}},  in {\em
  Automorphic Forms, Representation Theory and Arithmetic}, (Berlin,
  Heidelberg), pp.~303--355, Springer Berlin Heidelberg, 1981.
\newblock
  \url{https://people.mpim-bonn.mpg.de/zagier/files/scanned/EisensteinSelberg/fulltext.pdf}.

\bibitem{sarnakSpectralTheoryAutomorphic}
P.~Sarnak, {\it Spectral theory of automorphic forms},  {\em Lecture notes}.
  \url{https://web.math.princeton.edu/~gyujino/Sarnak_course.pdf}.

\bibitem{muhlenbruchTheoryMaassWave2020}
T.~Mühlenbruch and W.~Raji, {\em On the Theory of Maass Wave Forms}.
\newblock Springer, 2020.

\bibitem{Selberg}
A.~Selberg, {\it Harmonic analysis and discontinuous groups in weakly symmetric
  {Riemannian} spaces with applications to {Dirichlet} series},  {\em J. Indian
  Math. Soc., New Ser.} {\bf 20} (1956) 47--87.

\bibitem{hejhalSelbergTraceFormula1976}
D.~A. Hejhal, {\em The Selberg trace formula for PSL (2, R). 1}.
\newblock Lecture notes in mathematics. Springer, Berlin, 1976.

\bibitem{zbMATH04205458}
D.~H. Mayer, {\it The thermodynamic formalism approach to {Selberg}'s zeta
  function for {PSL}(2,{{\({\mathbb{Z}})\)}}},  {\em Bull. Am. Math. Soc., New
  Ser.} {\bf 25} (1991), no.~1 55--60.

\bibitem{Ruelle}
D.~Ruelle, {\it Zeta functions and statistical mechanics},  in {\em
  International conference on dynamical systems in mathematical physics},
  no.~40 in Ast\'erisque, pp.~167--176.
\newblock Soci\'et\'e math\'ematique de France, 1976.

\bibitem{FriedZeta}
D.~Fried, {\it The zeta functions of {Ruelle} and {Selberg}. {I}},  {\em Ann.
  Sci. {\'E}c. Norm. Sup{\'e}r. (4)} {\bf 19} (1986), no.~4 491--517.

\bibitem{sarnakSpectraHyperbolicSurfaces2003}
P.~Sarnak, {\it Spectra of hyperbolic surfaces},  {\em Bulletin of the American
  Mathematical Society} {\bf 40} (2003), no.~4 441–478.

\bibitem{Hejhal1991}
D.~A. Hejhal, {\em Eigenvalues of the Laplacian for PSL(2,Z): Some New Results
  and Computational Techniques}, pp.~59--102.
\newblock Springer Berlin Heidelberg, Berlin, Heidelberg, 1991.

\bibitem{luoQuantumErgodicityEigenfunctions1995}
W.~Luo and P.~Sarnak, {\it Quantum ergodicity of eigenfunctions on
  $\mathrm{PSL}_2(\mathbb{Z})\backslash\mathbb{H}$},  {\em Publications
  mathématiques de l’IHÉS} {\bf 81} (Dec., 1995) 207–237.
  \url{http://link.springer.com/10.1007/BF02699377}.

\bibitem{2005math6102F}
D.~W. {Farmer} and S.~{Lemurell}, {\it {Maass forms and their $L$-functions}},
  {\em arXiv Mathematics e-prints} (June, 2005) math/0506102,
  [\href{http://arxiv.org/abs/math/0506102}{{\tt math/0506102}}].

\bibitem{PhysRevLett.69.2188}
J.~Bolte, G.~Steil, and F.~Steiner, {\it Arithmetical chaos and violation of
  universality in energy level statistics},  {\em Phys. Rev. Lett.} {\bf 69}
  (Oct, 1992) 2188--2191.

\bibitem{laxScatteringTheoryAutomorphic1980}
P.~D. Lax and R.~S. Phillips, {\it Scattering theory for automorphic
  functions},  {\em Bulletin of the American Mathematical Society} {\bf 2}
  (1980), no.~2 261–295.

\bibitem{Gutzwiller_Jose_1990}
M.~C. Gutzwiller and J.~V. José, {\em Chaos in Classical and Quantum
  Mechanics}.
\newblock Jan, 1990.
\newblock \url{https://link.springer.com/book/10.1007/978-1-4612-0983-6}.

\bibitem{Castro:2011zq}
A.~Castro, M.~R. Gaberdiel, T.~Hartman, A.~Maloney, and R.~Volpato, {\it {The
  Gravity Dual of the Ising Model}},  {\em Phys. Rev. D} {\bf 85} (2012)
  024032, [\href{http://arxiv.org/abs/1111.1987}{{\tt arXiv:1111.1987}}].

\bibitem{Benjamin:2019stq}
N.~Benjamin, H.~Ooguri, S.-H. Shao, and Y.~Wang, {\it {Light-cone modular
  bootstrap and pure gravity}},  {\em Phys. Rev. D} {\bf 100} (2019), no.~6
  066029, [\href{http://arxiv.org/abs/1906.04184}{{\tt arXiv:1906.04184}}].

\bibitem{Maxfield:2020ale}
H.~Maxfield and G.~J. Turiaci, {\it {The path integral of 3D gravity near
  extremality; or, JT gravity with defects as a matrix integral}},  {\em JHEP}
  {\bf 01} (2021) 118, [\href{http://arxiv.org/abs/2006.11317}{{\tt
  arXiv:2006.11317}}].

\bibitem{Alday:2019vdr}
L.~F. Alday and J.-B. Bae, {\it {Rademacher Expansions and the Spectrum of 2d
  CFT}},  {\em JHEP} {\bf 11} (2020) 134,
  [\href{http://arxiv.org/abs/2001.00022}{{\tt arXiv:2001.00022}}].

\bibitem{Kaidi:2020ecu}
J.~Kaidi and E.~Perlmutter, {\it {Discreteness and integrality in Conformal
  Field Theory}},  {\em JHEP} {\bf 02} (2021) 064,
  [\href{http://arxiv.org/abs/2008.02190}{{\tt arXiv:2008.02190}}].

\bibitem{Benjamin:2020mfz}
N.~Benjamin, S.~Collier, and A.~Maloney, {\it {Pure Gravity and Conical
  Defects}},  {\em JHEP} {\bf 09} (2020) 034,
  [\href{http://arxiv.org/abs/2004.14428}{{\tt arXiv:2004.14428}}].

\bibitem{Meruliya:2021utr}
V.~Meruliya, S.~Mukhi, and P.~Singh, {\it {Poincar\'e Series, 3d Gravity and
  Averages of Rational CFT}},  {\em JHEP} {\bf 04} (2021) 267,
  [\href{http://arxiv.org/abs/2102.03136}{{\tt arXiv:2102.03136}}].

\bibitem{Mertens:2022ujr}
T.~G. Mertens, J.~Sim\'on, and G.~Wong, {\it {A proposal for 3d quantum gravity
  and its bulk factorization}},  {\em JHEP} {\bf 06} (2023) 134,
  [\href{http://arxiv.org/abs/2210.14196}{{\tt arXiv:2210.14196}}].

\bibitem{Chandra:2022bqq}
J.~Chandra, S.~Collier, T.~Hartman, and A.~Maloney, {\it {Semiclassical 3D
  gravity as an average of large-c CFTs}},  {\em JHEP} {\bf 12} (2022) 069,
  [\href{http://arxiv.org/abs/2203.06511}{{\tt arXiv:2203.06511}}].

\bibitem{DiUbaldo:2023hkc}
G.~Di~Ubaldo and E.~Perlmutter, {\it {AdS3 Pure Gravity and Stringy
  Unitarity}},  {\em Phys. Rev. Lett.} {\bf 132} (2024), no.~4 041602,
  [\href{http://arxiv.org/abs/2308.01787}{{\tt arXiv:2308.01787}}].

\bibitem{iwaniec2004analytic}
H.~Iwaniec and E.~Kowalski, {\em Analytic Number Theory}.
\newblock American Mathematical Society colloquium publications. American
  Mathematical Society, 2004.

\bibitem{GreenTao}
B.~Green and T.~Tao, {\it {The Möbius function is strongly orthogonal to
  nilsequences}},  {\em Annals of Mathematics} {\bf 175} (2012), no.~2
  541--566. \url{https://arxiv.org/abs/0807.1736}.

\bibitem{Kaisa}
K.~Matomaki and M.~Radziwill, {\it Multiplicative functions in short
  intervals},  {\em Annals of Mathematics} {\bf 183} (01, 2015).
  \url{https://arxiv.org/abs/1501.04585}.

\bibitem{TaoBlogPost}
T.~Tao, {\it {The Chowla conjecture and the Sarnak conjecture (blog post)}},
  2012.
\newblock
  \url{https://terrytao.wordpress.com/2012/10/14/the-chowla-conjecture-and-the-sarnak-conjecture/}.

\bibitem{sarnakThreeLecturesMobius}
P.~Sarnak, ``{Three Lectures on the Möbius Function Randomness and
  Dynamics}.''
\newblock
  \url{https://publications.ias.edu/sites/default/files/MobiusFunctionsLectures(2).pdf}.

\bibitem{Eberhardt:2023mrq}
L.~Eberhardt, {\it {Notes on crossing transformations of Virasoro conformal
  blocks}},  \href{http://arxiv.org/abs/2309.11540}{{\tt arXiv:2309.11540}}.

\bibitem{Eberhardt:2022wlc}
L.~Eberhardt, {\it {Off-shell Partition Functions in 3d Gravity}},  {\em
  Commun. Math. Phys.} {\bf 405} (2024), no.~3 76,
  [\href{http://arxiv.org/abs/2204.09789}{{\tt arXiv:2204.09789}}].

\bibitem{Lewis1997PeriodFA}
J.~S. Lewis and D.~Zagier, {\it {Period functions and the Selberg zeta function
  for the modular group.}},  1997.
\newblock \url{https://api.semanticscholar.org/CorpusID:14657049}.

\bibitem{ZagierLewisMaaass}
{\it {Period functions for Maass wave forms. I}},  {\em Annals of Mathematics
  153 (2001) 191-258}.
  \url{https://people.mpim-bonn.mpg.de/zagier/files/doi/10.2307/2661374/fulltext.pdf}.

\bibitem{ZagierSelbergZeta}
{\it {New points of view on the Selberg zeta function}},  {\em Proceedings of
  the Japanese-German Seminar "Explicit Structures of Modular Forms and Zeta
  Functions", Ryushi-do (2002), 1-10}.
  \url{https://people.mpim-bonn.mpg.de/zagier/files/tex/NewPointsSelbergZeta/fulltext.pdf}.

\bibitem{urgenFischer}
J.~Fischer, {\it {An approach to the Selberg trace formula via the Selberg
  zeta-function}},  {\em Lecture Notes in Math} {\bf 1253}.

\bibitem{eiseleDynamicalZetaFunctions1994}
M.~Eisele and D.~Mayer, {\it {Dynamical zeta functions for Artin’s billiard
  and the Venkov--Zograf factorization formula}},  {\em Physica D: Nonlinear
  Phenomena} {\bf 70} (Feb., 1994) 342–356.
  \url{http://arxiv.org/abs/chao-dyn/9307001}.

\bibitem{Maldacena:2004rf}
J.~M. Maldacena and L.~Maoz, {\it {Wormholes in AdS}},  {\em JHEP} {\bf 02}
  (2004) 053, [\href{http://arxiv.org/abs/hep-th/0401024}{{\tt
  hep-th/0401024}}].

\bibitem{Cotler:2020ugk}
J.~Cotler and K.~Jensen, {\it {AdS$_{3}$ gravity and random CFT}},  {\em JHEP}
  {\bf 04} (2021) 033, [\href{http://arxiv.org/abs/2006.08648}{{\tt
  arXiv:2006.08648}}].

\bibitem{Yan:2023rjh}
C.~Yan, {\it {More on torus wormholes in 3d gravity}},  {\em JHEP} {\bf 11}
  (2023) 039, [\href{http://arxiv.org/abs/2305.10494}{{\tt arXiv:2305.10494}}].

\bibitem{Penington:2019kki}
G.~Penington, S.~H. Shenker, D.~Stanford, and Z.~Yang, {\it {Replica wormholes
  and the black hole interior}},  {\em JHEP} {\bf 03} (2022) 205,
  [\href{http://arxiv.org/abs/1911.11977}{{\tt arXiv:1911.11977}}].

\bibitem{McNamara:2020uza}
J.~McNamara and C.~Vafa, {\it {Baby Universes, Holography, and the Swampland}},
   \href{http://arxiv.org/abs/2004.06738}{{\tt arXiv:2004.06738}}.

\bibitem{Benini:2022hzx}
F.~Benini, C.~Copetti, and L.~Di~Pietro, {\it {Factorization and global
  symmetries in holography}},  {\em SciPost Phys.} {\bf 14} (2023), no.~2 019,
  [\href{http://arxiv.org/abs/2203.09537}{{\tt arXiv:2203.09537}}].

\bibitem{Usatyuk:2024isz}
M.~Usatyuk and Y.~Zhao, {\it {Closed universes, factorization, and ensemble
  averaging}},  \href{http://arxiv.org/abs/2403.13047}{{\tt arXiv:2403.13047}}.

\bibitem{Witten:2023qsv}
E.~Witten, {\it {Algebras, regions, and observers}},  {\em Proc. Symp. Pure
  Math.} {\bf 107} (2024) 247--276,
  [\href{http://arxiv.org/abs/2303.02837}{{\tt arXiv:2303.02837}}].

\bibitem{Witten:2023xze}
E.~Witten, {\it {A background-independent algebra in quantum gravity}},  {\em
  JHEP} {\bf 03} (2024) 077, [\href{http://arxiv.org/abs/2308.03663}{{\tt
  arXiv:2308.03663}}].

\bibitem{Taylor:2018xcy}
M.~Taylor, {\it {$T \bar{T}$ deformations in general dimensions}},  {\em Adv.
  Theor. Math. Phys.} {\bf 27} (2023), no.~1 37--63,
  [\href{http://arxiv.org/abs/1805.10287}{{\tt arXiv:1805.10287}}].

\bibitem{Hartman:2018tkw}
T.~Hartman, J.~Kruthoff, E.~Shaghoulian, and A.~Tajdini, {\it {Holography at
  finite cutoff with a $T^2$ deformation}},  {\em JHEP} {\bf 03} (2019) 004,
  [\href{http://arxiv.org/abs/1807.11401}{{\tt arXiv:1807.11401}}].

\bibitem{jacquet1970automorphic}
H.~Jacquet and R.~Langlands, {\em Automorphic Forms on GL (2)}.
\newblock Springer-Verlag, 1970.
\newblock \url{https://link.springer.com/book/10.1007/BFb0058988}.

\bibitem{Bump_1997}
D.~Bump, {\em Automorphic Forms and Representations}.
\newblock Cambridge Studies in Advanced Mathematics. Cambridge University
  Press, 1997.

\end{thebibliography}\endgroup

\end{document}